\documentclass{jsdarticle}
\usepackage{amssymb}
\usepackage{xspace}
\usepackage{color}
\usepackage{wasysym}
\usepackage{tikz}
\usepackage{siunitx}
\newcommand{\nc}{\newcommand}
\nc{\rnc}{\renewcommand}
\nc{\bs}{\boldsymbol}
\nc{\real}{\operatorname{Re}}
\nc{\imag}{\operatorname{Im}}
\nc{\etal}{et al.\xspace}
\nc{\ie}{i.\,e.\xspace}
\nc{\eg}{e.\,g.\xspace}
\nc{\cf}{cf.\xspace}
\nc{\fp}{\,.}
\nc{\fk}{\,,}
\nc{\mrm}[1]{\mathrm{#1}} 
\nc{\dd}{\mathrm{d}}
\nc{\ee}{\mrm{e}}
\nc{\ii}{\mrm{i}}
\nc{\tra}{^{\mathrm T}}
\nc{\BYUUW}{{\sf{BYU-UW}}\xspace}
\nc{\ETH}{{\sf{ETH}}\xspace}
\nc{\FAU}{{\sf{FAU}}\xspace}
\nc{\ICL}{{\sf{ICL}}\xspace}
\nc{\NWPU}{{\sf{NWPU}}\xspace}
\nc{\SNL}{{\sf{SNL}}\xspace}
\nc{\SU}{{\sf{SU}}\xspace}
\nc{\USTUTT}{{\sf{USTUTT}}\xspace}
\nc{\aref}[1]{\ref{asec:#1}}
\nc{\sref}[1]{Sect.~\ref{sec:#1}}
\nc{\srefs}[1]{Sect.~\ref{sec:#1}}
\nc{\srefo}[1]{\ref{sec:#1}}
\nc{\ssref}[1]{Subsect.~\ref{sec:#1}}
\nc{\ssrefs}[1]{Subsect.~\ref{sec:#1}}
\nc{\ssrefo}[1]{\ref{sec:#1}}
\nc{\eref}[1]{Eq.~\ref{eq:#1}}
\nc{\erefs}[1]{Eqs.~\ref{eq:#1}}
\nc{\erefo}[1]{\ref{eq:#1}}
\nc{\fref}[1]{Fig.~\ref{fig:#1}}
\nc{\frefs}[1]{Figs.~\ref{fig:#1}}
\nc{\frefo}[1]{\ref{fig:#1}}
\nc{\tref}[1]{Tab.~\ref{tab:#1}}
\nc{\fig}[4][tbh]{
\begin{figure}[#1]
\centering
\includegraphics[width=#4\textwidth]{figs/#2}
\caption{#3\label{fig:#2}}
\end{figure}}
\nc{\ea}[1]{
\begin{eqnarray}
#1\end{eqnarray}}
\nc{\COMMENT}[1]{\textcolor{red}{#1}}
\nc{\mm}{\bs}
\graphicspath{{./figs/}}
\PassOptionsToPackage{hyphens}{url}\usepackage{hyperref}
\begin{document}

\title{Experimental analysis of the TRC benchmark system}

\author[1]{Arati Bhattu}
\author[2]{Svenja Hermann}
\author[3]{Nidhal Jamia}
\author[4]{Florian Müller}
\author[4]{Maren Scheel}
\author[5]{Christoph Schwingshackl}
\author[6]{H. Nevzat Özgüven}
\author[4]{Malte Krack \email{krack@ila.uni-stuttgart.de}}

\affil[1]{Rice University Houston, USA}
\affil[2]{FEMTO-ST Institute Besançon, France}
\affil[3]{Swansea University, UK}
\affil[4]{University of Stuttgart, Germany}
\affil[5]{Imperial College London, UK}
\affil[6]{Middle East Technical University, Ankara, Turkey}

\begin{abstract}
The Tribomechadynamics Research Challenge (TRC) was a blind prediction of the vibration behavior of a thin plate clamped on two sides using bolted joints.
Specifically, the natural frequency and damping ratio of the fundamental bending mode were requested as function of the amplitude, starting from the linear regime until high levels, where both frictional contact and nonlinear bending-stretching coupling become relevant.
The predictions were confronted with experimental results in a companion paper; the present article addresses the experimental analysis of this benchmark system.
Amplitude-dependent modal data was obtained from phase resonance and response controlled tests.
An original variant of response controlled testing is proposed:
Instead of a fixed frequency interval, a fixed phase interval is analyzed.
This way, the high excitation levels required outside resonance, which could activate unwanted exciter nonlinearity, are avoided.
The consistency of the nonlinear modal testing methods, with each other, and with conventional linear tests at low amplitudes, is carefully analyzed.
Comparisons of nonlinear-mode based predictions with direct frequency response curve measurements (at fixed excitation level) serve as additional cross-validation.
Overall, these measures have permitted to gain high confidence in the acquired modal data.
The different sources of the remaining uncertainty were further analyzed.
A low reassembly-variability but a moderate time-variability were identified, where the latter is attributed to some thermal sensitivity of the system.
Two nominally identical plates were analyzed, which both have an appreciable initial curvature, and a significant effect on the vibration behavior was found depending on whether the plate is aligned/misaligned with the support structure.
Further, a 1:2 nonlinear modal interaction with the first torsion mode was observed, which only occurs in the aligned configurations.
\keywords{friction damping; jointed structures; geometric nonlinearity; nonlinear dynamics; nonlinear modal analysis}
\end{abstract}

\maketitle

\section{Introduction\label{sec:intro}}
The benchmark system of the Tribomechadynamics Research Challenge (TRC) is depicted in \fref{02_TRC_Challenge_CAD}.
It consists of four main parts: a thin plate (\emph{panel}), a support (monolithic piece comprising two pillars and a thick rear plate), and two blades.
The panel is bolted between blades and pillars with six M6 bolts and washers per side spaced based on industry recommendations.
The rear plate is to be bolted on the slip table of a large shaker.
The panel's nominal initial geometry is flat.
The contact surfaces with the pillars are flat, but deliberately aligned under a nominal angle of $2.2^\circ$ about the $z$-axis ($1.1^\circ$ inclination on each side), so that the panel is arched in the bolted configuration.
The CAD models, the technical drawings and the design documentation (including assembly instructions) are available in a data repository~\cite{Muller.2022c}.
\fig[htb]{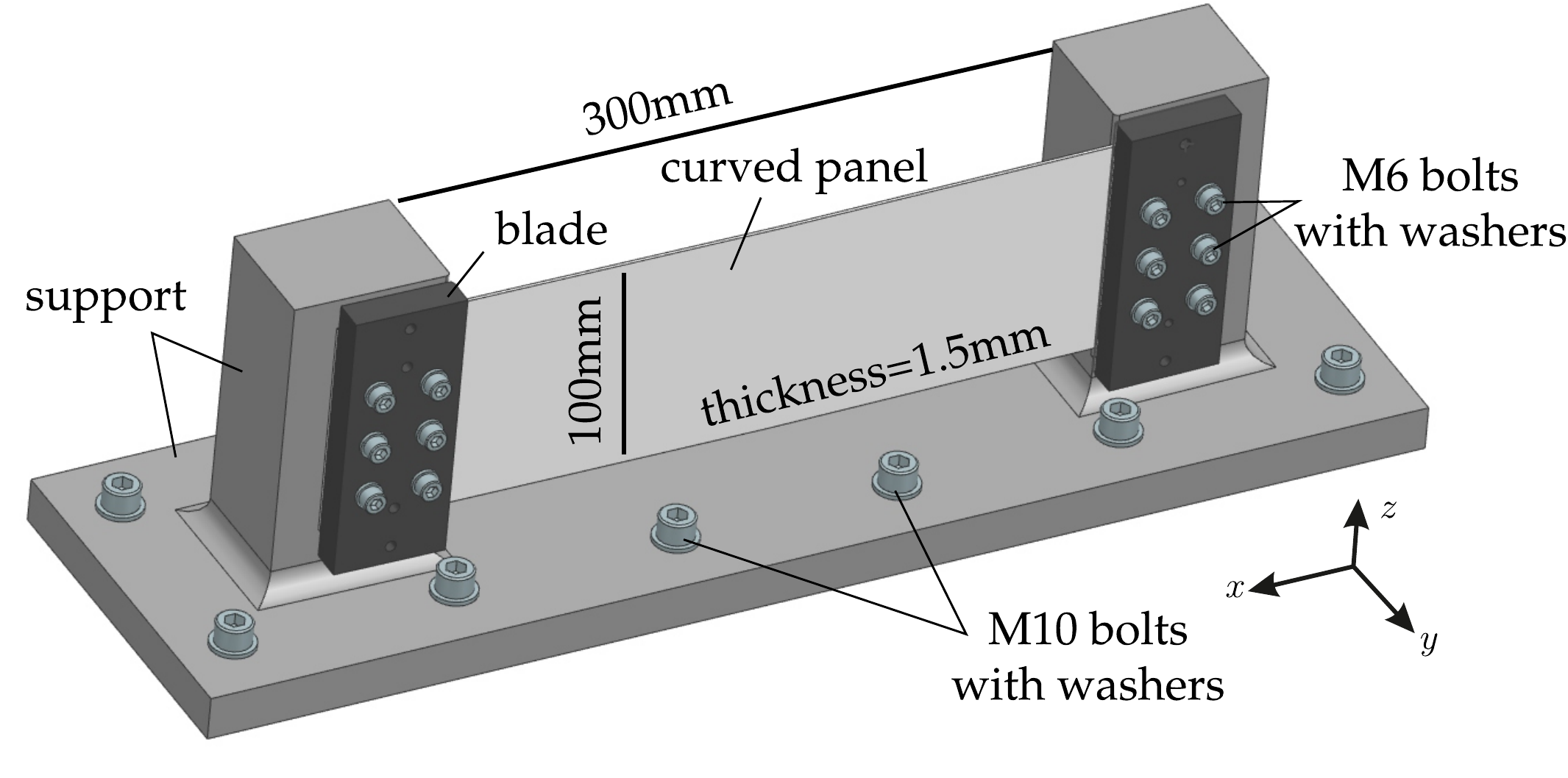}{TRC Benchmark system}{0.8}
\\
Thin panels have a widespread use in aircraft, space, and wind turbine industries to achieve high strength-to-weight ratios.
A key feature of these systems is that both frictional contact and geometric nonlinearity are relevant.
More specifically, panels are commonly assembled via mechanical joints using fasteners (\eg pins, rivets, bolts), and dry frictional and unilateral interactions occur at the contact interfaces.
On the other hand, due to the clamped ends, bending induces membrane stretching/compression, which, in turn, affects the bending stiffness; this bending-stretching coupling is an important example of geometric nonlinearity.
The individual modeling of those nonlinearities and the prediction of their effects on the vibration behavior is not a trivial task, as outlined in the following.
\\
At least in the case of linear-elastic material behavior, the modeling of geometrically nonlinear structures is well-established, and its effects on the vibration behavior are well-understood, see \eg \cite{nayf1979,Thomsen.2003,Thomas.2016}.
In particular, bending-stretching coupling can have a severe effect on the natural frequencies, and trigger nonlinear modal interactions already for bending deformations in the order of magnitude of the panel thickness.
The derivation of accurate and efficient reduced-order models for analysis and design purposes is still an active field of research \cite{Mignolet.2013,Touze.2021,Vizzaccaro.2022}.
Moreover, predictive modeling is hampered by the high sensitivity to the boundary stiffness \cite{Ibrahim.2005}, thermal effects \cite{HARRAS.2002,Jain.2020}, and geometric imperfections/shape defects \cite{Marconi.2020,Marconi.2021,Saccani.2023}.
\\
It is well-known that dissipative dry frictional interactions in the mechanical joints are often the main source of mechanical damping in built-up structures \cite{Gaul1997,popp2003a}.
Compared to geometric nonlinearity, the physics of frictional contact of assembled structures is less well-understood, and phenomenological models have to be used.
Predictive modeling is further hampered by a high sensitivity to the local material behavior and the contact surface topography, which are uncertain at the design stage, and vary with the thermo-mechanical load history due to wear and corrosion \cite{Brake.2018}.
\\
Remarkably, geometric and contact nonlinearity is rarely considered simultaneously.
A mutual interaction is to be expected:
On the one hand, the extent of bending-stretching coupling depends on the effective axial support stiffness, which is highly dependent on the contact interactions and, thus, amplitude-dependent.
On the other hand, one can expect that bending deformation leads to opening contacts, whereas at higher amplitudes, stretching should increase the tangential load on the contact interfaces.
To the best of the authors' knowledge, the study of Yun and Bachau \cite{Yun.1998} is the only experimental work on panel assemblies that considered both nonlinearities.
In particular, they attribute a $12~\%$ natural frequency drop with increasing amplitude to the contact interactions in the joints, and a subsequent $4\%$ increase to the geometric nonlinearity.
The development of prediction methods for thin-walled structures with contact constraints is an important area of active research.
The prediction approaches pursued by the participating teams of the TRC are described in \cite{TRCprediction}, and compared to the experimental results obtained with the methods described in the present work.
\\
The experimental identification of the amplitude-dependent modal properties of the benchmark system is not a trivial task.
As in the linear case, impact hammer modal testing is popular also in the nonlinear case \cite{Deaner.2015,Kuether.2016,Jin.2019}.
However, this drives energy into many modes, which then interact in a way that cannot be simply deconvoluted.
This limits impact hammer modal testing to the very weakly nonlinear regime \cite{Kuether.2016}.
Further, for the given benchmark system, applying an impact at a singular location on the panel is expected to either introduce local damage or to not excite sufficiently high vibrations ($3~\mrm{mm}$ amplitude at the panel center were requested in the challenge).
Thus, shaker-based testing has to be used.
For nonlinear vibration testing, shaker-stinger excitation is often used.
The large displacement amplitudes of the panel, again, make this excitation strategy practically impossible:
For such large amplitudes, distortions due to strongly nonlinear exciter behavior, as well as intrusiveness due to parasitic stiffness and/or mass contributions are to be expected. 
For those reasons, the benchmark system was designed to be tested on a slip table of a large shaker.
Base excitation belongs to the most popular forms of load application for vibration testing, especially in aerospace engineering \cite{Beliveau.1986,ewin1995,Worden.2001}.
As the structure under test is mounted on the excitation system, there is no practical way to remove the excitation in order to initiate a free decay \cite{Feldman.1997,Peeters.2011,Dion.2013b,Londono.2015}.
It thus becomes crucial to carefully design the test in such a way that one identifies the dynamics of the structure under test only (not including the dynamics of the excitation system) \cite{Tomlinson.1979,McConnell.1995,Krack.2021}.
Here and in the following, the base acceleration is considered as excitation signal, and a focus is placed on the steady-state response to sinusoidal excitation.
\\
To characterize the nonlinear vibration behavior around a particular resonance, one should cover some parts of the frequency-response surface spanned by the excitation frequency, the response amplitude and the excitation amplitude. 
The most common methods are:
\begin{enumerate}
    \item Phase Resonance Testing (PRT),
    \item Response Controlled Testing (RCT), and
    \item Excitation Controlled Testing (ECT).
\end{enumerate}
During PRT, the phase lag between response and excitation is fixed, and the excitation or response level is step-wise in-/decreased.
During RCT, the response level is kept fixed, the excitation frequency is step-wise in-/decreased, and this procedure is repeated at different response levels.
During ECT, the excitation level is kept fixed, the excitation frequency is step-wise in-/decreased, and this procedure is repeated at different excitation levels.
All those methods have the strength that they are not limited to a particular spatial location/distribution or mathematical form of the nonlinear behavior, which is an important advantage over many nonlinear system identification techniques \cite{Kerschen.2006}.
Among the three techniques, PRT leads to the smallest amount of response points (only one per response/excitation level).
The amplitude-dependent modal frequency is directly tracked (\emph{backbone}), and the modal damping ratio can be obtained from the balance of dissipated and supplied power in period-average \cite{Scheel.2018,Muller.2022}.
RCT has the important advantage that it leads to quasi-linear behavior, so that standard methods from linear vibration theory can be used to estimate the frequency response functions for each response level, and to identify the modal properties \cite{Arslan.2008,Link.2011,Karaagacl.2021,Karaagacl.2022,Koyuncu.2022}.
In contrast, ECT generally leads to nonlinear phenomena, possibly including coexisting limit states for the same frequency and jump phenomena.
To recover local uniqueness of the limit state, it is common to control the phase lag instead of the excitation frequency.
Phase control was also needed to recover coexisting limit states for the considered benchmark system.
The phase-controlled variant of ECT will be simply referred to as ECT henceforth.
Amplitude-dependent modal properties can be obtained by regression of a single-nonlinear-mode model \cite{seti1992,gibe2003}.
For completeness, one should mention the acquisition of so-called S-curves as an additional variant beyond the three aforementioned ones.
Here, the excitation frequency is fixed and the response level is step-wise in-/decreased, and this is repeated for different excitation frequencies.
This is commonly done in conjunction with Control-Based Continuation \cite{Renson.2016b,Renson.2017b}.
\\
A key element of both PRT and (phase-controlled) ECT is a means to achieve a certain phase lag between response and excitation.
As the limit state can be unstable in open-loop conditions (\eg overhanging branch of frequency-response curve), feedback control of the phase must be used.
This is most commonly achieved with a phase-locked loop, which is very well-known in electrical and control engineering.
In the context of nonlinear vibration testing, first applications date back at least to the 1970s \cite{Axelsson.1976}, and it has become popular more recently \cite{Denis.2018,Scheel.2018,Scheel.2020b}.
It should be remarked that such a controller is not yet available as a commercial vibration testing tool, and so far a rather heuristic tuning of the control parameters is needed.
This is in contrast to RCT, which can be carried out with commercial hard- and software \cite{Karaagacl.2021}.
\\
Compared to linear vibration analysis, empirical evidence is still scarce for the aforementioned nonlinear methods.
No publication is known to date where RCT is implemented using base excitation.
It is noteworthy that ECT was used for cross-validation of modal data obtained from PRT \cite{Peter.2018,Scheel.2018,Scheel.2020b,Muller.2022} or from RCT \cite{Karaagacl.2021}.
Further, consistency between PRT and Control-Based Continuation was demonstrated in \cite{Abeloos.2022} for one particular test case.
\\
One objective of the present work is to assess the current state of the art of nonlinear testing methods for a challenging benchmark system.
Also, an original testing approach is proposed, phase-controlled RCT.
Using these techniques, the nonlinear modal properties of the TRC benchmark system are obtained, and their variability is quantified and characterized.
This is a crucial prerequisite for the further development of appropriate vibration prediction approaches.
Finally, some interesting insight into the role and interplay of bending-stretching coupling and frictional contact is obtained, which leads to strongly nonlinear behavior, including modal interactions.
The experimental methodology is described in \sref{method}.
The results are discussed in \sref{results}.
Conclusions are drawn in \sref{concl}.

\section{Methodology\label{sec:method}}
As mentioned before, the benchmark system was mounted on the slip table of a large shaker.
Setup and instrumentation are described in \ssref{setup}.
Conventional linear modal analysis was carried out by applying low-level broadband excitation via the shaker (\ssref{linear}).
The nonlinear tests, applied to identify the amplitude-dependent modal frequency and damping ratio of the fundamental bending mode, are described in \ssref{PRT}-\ssrefo{RCT}.
For cross-validation, frequency responses were also acquired using ECT as described in \ssref{ECT}.
The test program, which was carried out for both sides of the two nominally identical panels is described in \ssref{protocol}.

\subsection{Setup and instrumentation\label{sec:setup}}
Setup and instrumentation are illustrated in \fref{testrig}.
The benchmark system was subjected to base excitation.
The velocity at a point on the right blade was measured using a single laser-Doppler vibrometer (LDV).
Preliminary tests confirmed that the bottom plate, the pillars and the blades move in very good approximation as a rigid body in one direction only, so that the given velocity can be used as reference for the base motion.
The velocity at the panel center, relative to the base, was measured using a differential LDV.
This response velocity, along with the base velocity, was used for the control tasks described in \ssrefs{PRT}-\ssrefo{ECT}.
The control tasks were implemented using a dSPACE MicoLabBox, which operates at a sampling frequency of $10~\mrm{kHz}$.
A multi-point vibrometer (MPV) was used to measure the velocity at 15 points positioned in a 5 $\times$ 3 grid on the panel (\fref{MPVlocation}).
Reflection tape was applied at all sensor locations to improve the signal-to-noise ratio.
\begin{figure}[htb]
     \begin{subfigure}[c]{0.44\textwidth}
         \centering
         \includegraphics[width=\textwidth]{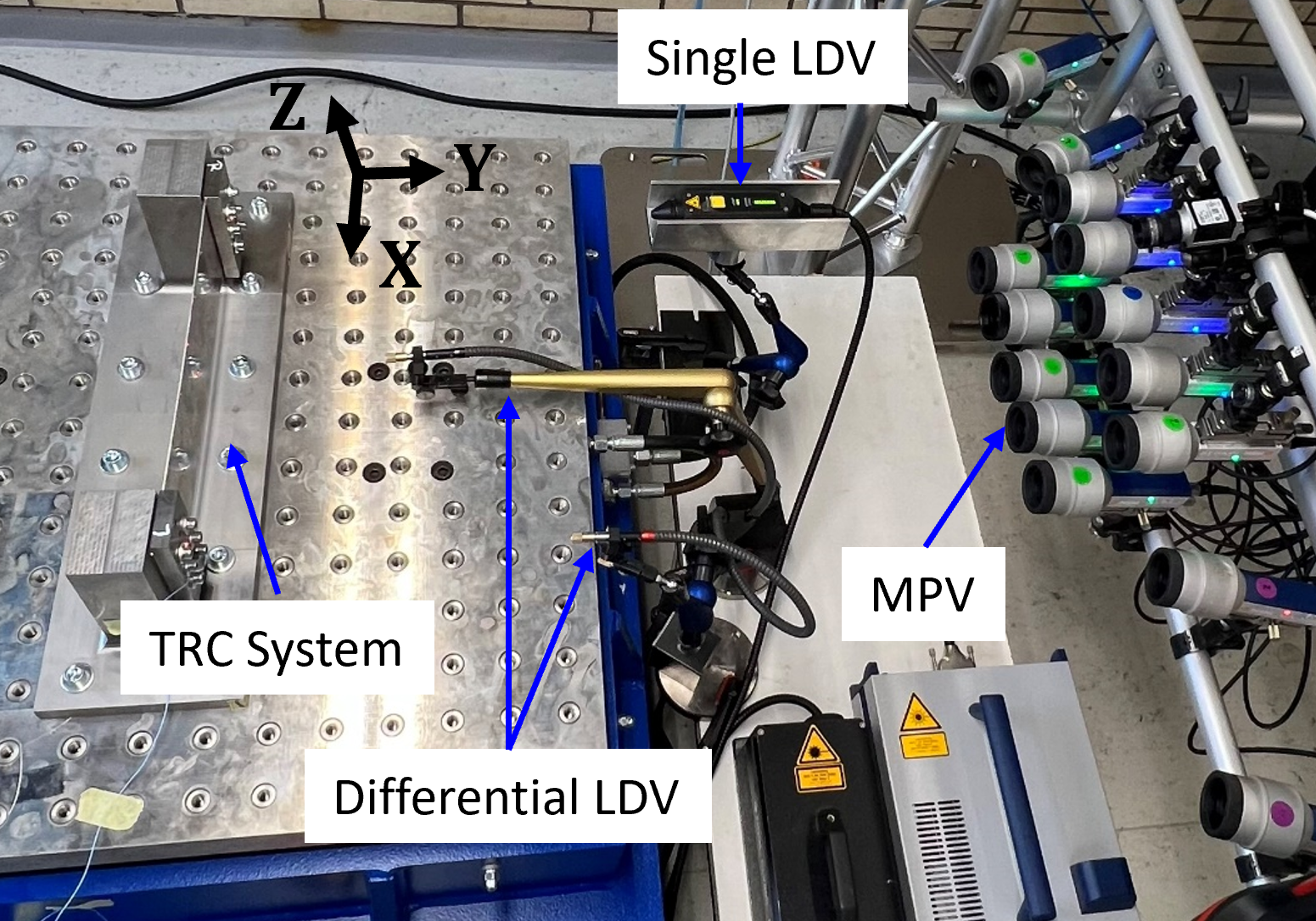}
         \caption{Instrumentation}
         \label{fig:setup2}
     \end{subfigure}
     \begin{subfigure}[c]{0.55\textwidth}
         \centering
         \includegraphics[width=\textwidth]{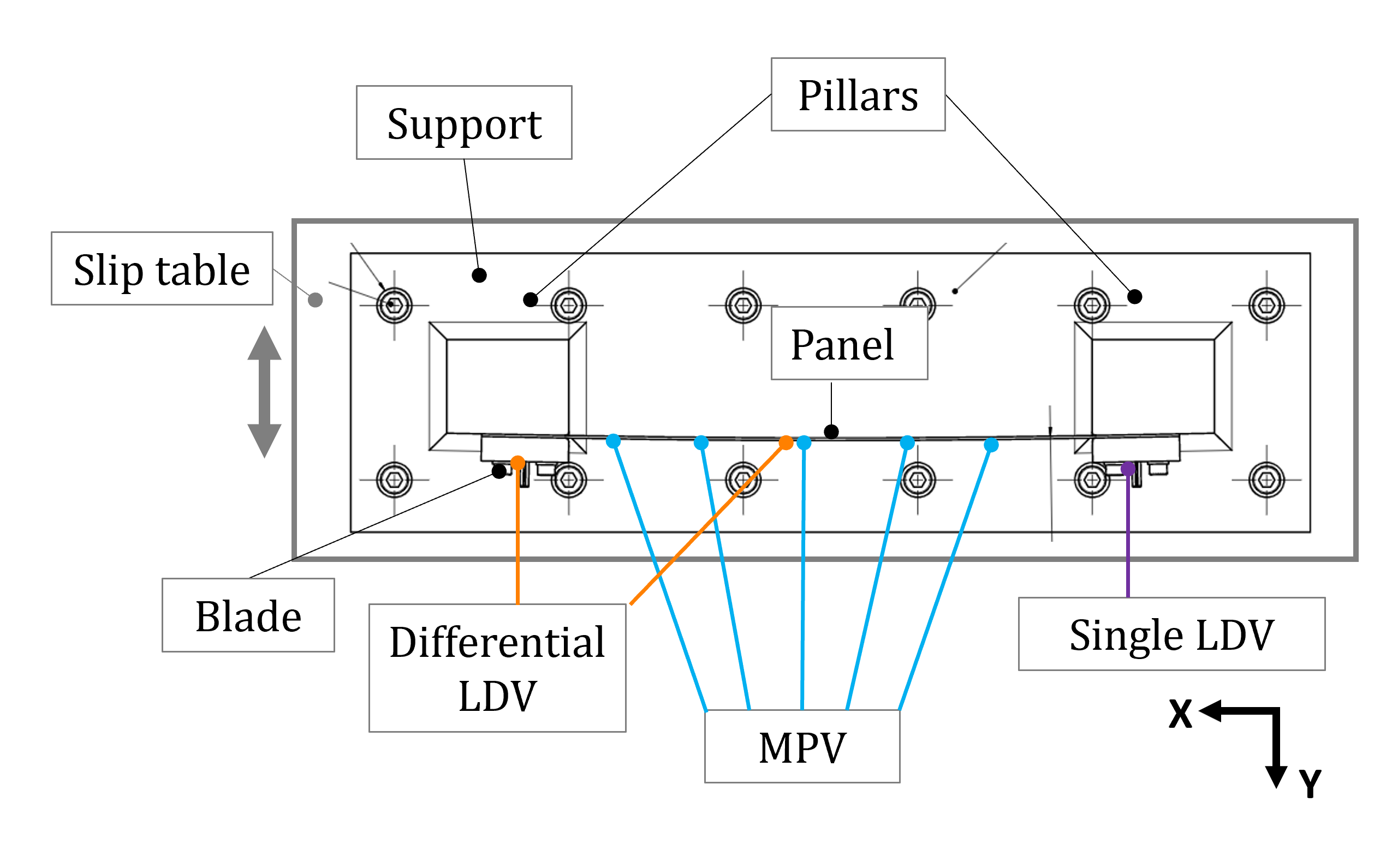}
         \caption{Schematic top view}
         \label{fig:setup3}
     \end{subfigure}
     \caption{Experimental test rig used during the Tribomechadynamics Research Challenge 2022.
     }
        \label{fig:testrig}
\end{figure}
\begin{figure}[htb]
     \centering
         \includegraphics[width=0.8\textwidth]{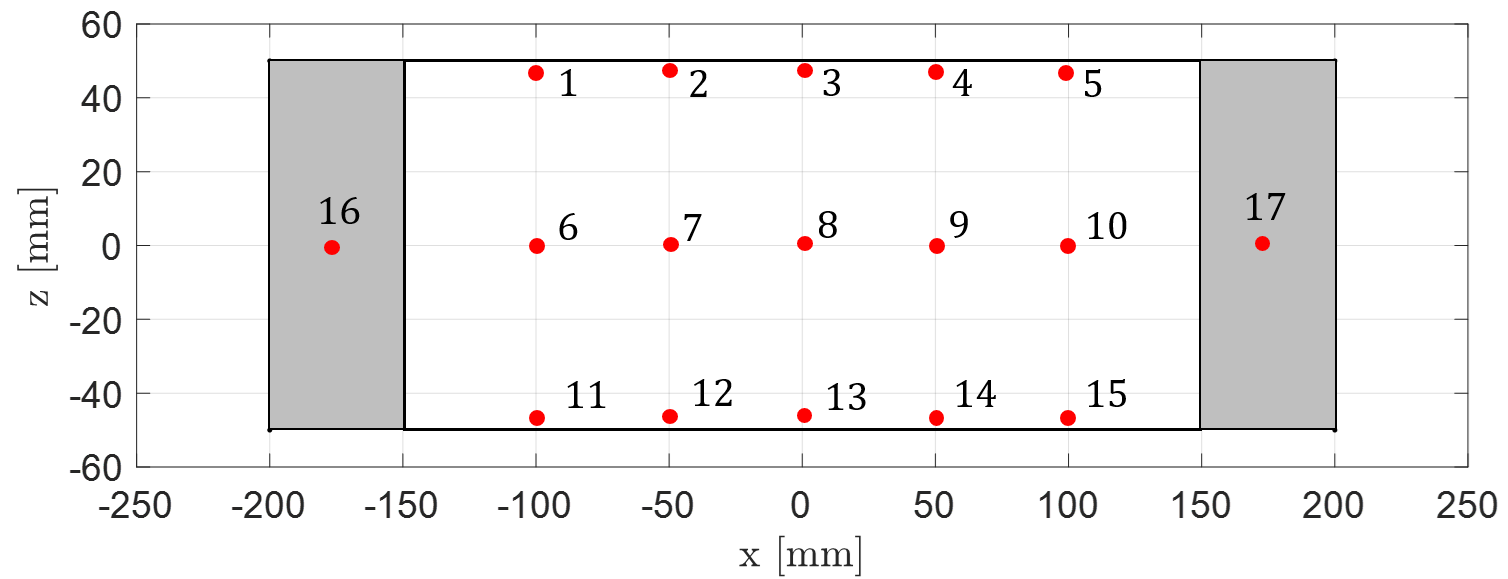}
         \caption{MPV sensor locations.}
         \label{fig:MPVlocation}
\end{figure}

\subsection{Linear modal test\label{sec:linear}}
For the linear modal analysis, a pseudo-random broadband signal was generated and fed to the shaker amplifier.
The frequency band was wide enough to ensure a good estimation of the fundamental modal frequencies, and the excitation level was small enough to avoid the activation of significant nonlinearity.
As the fundamental bending mode is well-separated, the simple peak-picking method was deemed sufficient to identify natural frequency and modal damping ratio.

\subsection{Phase Resonance Test (PRT)\label{sec:PRT}}
\begin{figure}[htb]
     \centering
     \begin{subfigure}[c]{0.42\textwidth}
         \centering
     \begin{minipage}{\textwidth}
        \includegraphics[width=1\textwidth]{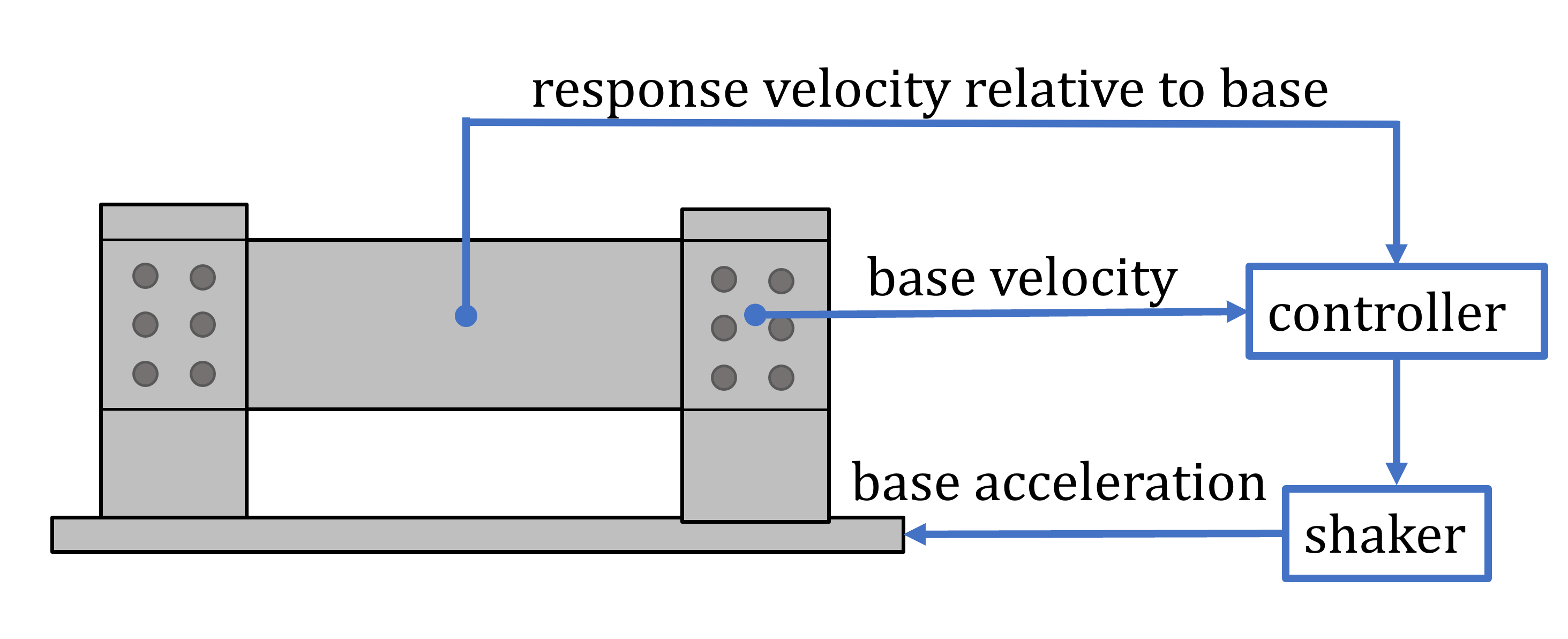}
        \end{minipage}
     \end{subfigure}
     \hfill
     \begin{subfigure}[c]{0.27\textwidth}
         \centering
     \begin{minipage}{\textwidth}
        \includegraphics[width=1\textwidth]{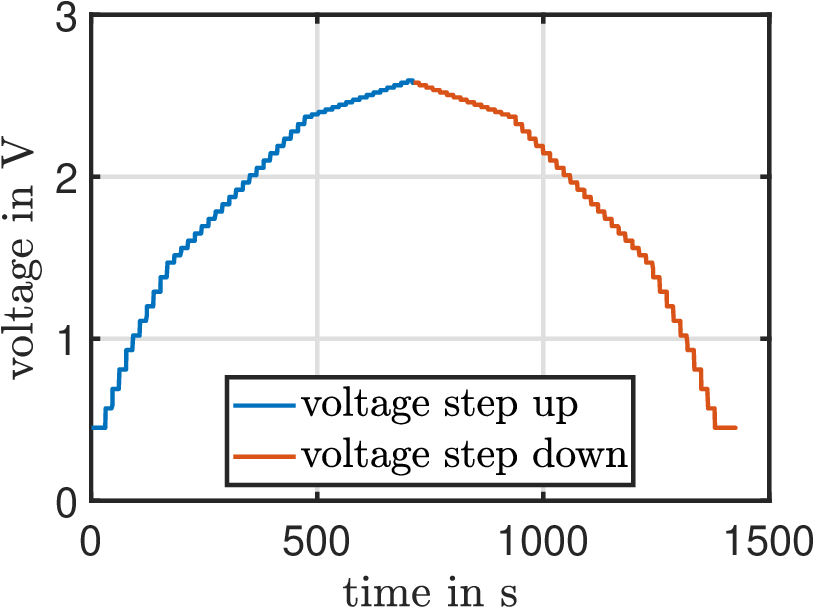}
        \end{minipage}
     \end{subfigure}
     \hfill
     \begin{subfigure}[c]{0.27\textwidth}
         \centering
     \begin{minipage}{\textwidth}
        \includegraphics[width=1\textwidth]{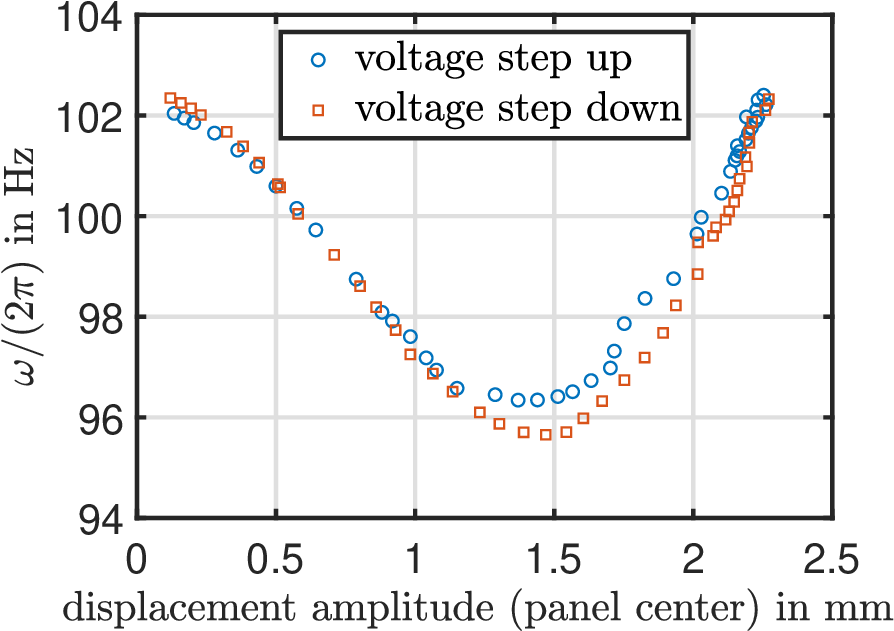}
        \end{minipage}
     \end{subfigure}
     \caption{Feedback control loop for nonlinear testing (left), step-wise in- and then decreasing voltage used for PRT (middle), resulting modal frequency vs. response level (right).}
	\label{fig:PLL1}
\end{figure}
%
Three types of nonlinear tests were carried out, a \emph{Phase Resonance Test} (PRT), a \emph{Response Controlled Test} (RCT), and an \emph{Excitation Controlled Test} (ECT).
In all these tests, the feedback control loop illustrated in \fref{PLL1}-left was used.
The base velocity was used as excitation signal, and the velocity at the panel center, relative to the base, was used as response signal.
In the case of the PRT, the control target is to reach phase resonance; \ie, the response velocity shall lag $90^\circ$ behind the base velocity.
This applies only to the fundamental harmonic component of those signals.
Throughout the nonlinear tests presented in this work, a purely sinusoidal voltage input was applied to the shaker amplifier.
To reach phase resonance, a phase-locked loop was employed.
Here, the phase of base and response velocity was estimated using synchronous demodulation.
The control error is the deviation between the estimated phase difference and the set value of $90^\circ$.
A proportional-integral controller was used.
The output of this controller is the instantaneous frequency, which is integrated to obtain the instantaneous phase of the voltage.
The cosine of this phase is taken, which is then multiplied by a prescribed voltage amplitude.
The resulting signal is fed to the shaker amplifier.
The voltage amplitude is step-wise increased and then decreased as shown in \fref{PLL1}-middle.
Each voltage amplitude was maintained for $16~\mrm{s}$, which corresponds to about 1500-1800 periods of the fundamental mode.
Indeed, it was observed that the transients have sufficiently decayed in that time span, and the recorded time span was found sufficiently long to obtain stabilized results, throughout the PRT results reported in this work.
The remaining phase error was $<2^\circ$, and the instantaneous frequency had a standard deviation $<0.35~\mrm{Hz}$.
\\
The mean of the instantaneous frequency (output of phase-locked loop) was adopted as modal frequency.
Exemplary results are shown in \fref{PLL1}-right.
The periodic modal vibration was acquired with the MPV.
The modal damping ratio was obtained by considering the power balance between dissipation and power supplied by distributed inertia forces, in period-average.
The latter was estimated using the model-free variant of the method proposed in \cite{Muller.2022}, relying on the velocity acquired at the different points indicated in \fref{MPVlocation}.
Here, again, the balance was restricted to the (controlled) fundamental harmonic component, as explained in \cite{Muller.2022}.

\subsection{Response Controlled Test (RCT)\label{sec:RCT}}
As stated in the introduction, RCT entails applying sinusoidal excitation where the level is adjusted to maintain a specified response amplitude around resonance.
This way, one obtains quasi-linear frequency response functions, so that conventional techniques for the identification of the modal properties can be used.
By repeating this for different target response amplitudes, one obtains amplitude-dependent modal properties.
To the best of the authors' knowledge, RCT has so far not been applied to the case of base excitation.
Conceptually it is not very difficult to adjust the method accordingly, as shown in the following.
\\
As in \cite{Karaagacl.2021}, it is assumed that the vibration response is dominated by a single nonlinear mode.
Under steady-state conditions, the modal oscillator equation takes the form \cite{szem1979,krac2013a,Krack.2021}
\ea{
\left(~-\Omega^2+2 \,D\left(a\right) \,\omega\left(a\right) \,\ii \Omega +\omega^2\left(a\right)~\right)\, a e^{\ii\, \theta}={\mm \varphi}_1^{\mrm H}\left(a\right) \hat{\mm f}_1 \fp \label{eq:SNMT}
}
Herein, $\Omega$ is the fundamental (angular) oscillation frequency, which is assumed to be constant and imposed here by the external forcing $\mm f$, with fundamental complex Fourier coefficient $\hat{\mm f}_1$, and $\square^{\mrm{H}}$ denotes the complex-conjugate transpose (Hermitian).
$\omega$ and $D$ are the modal (angular) frequency and damping ratio, and $\mm\varphi_1$ is the fundamental harmonic of the modal deflection shape, all of which depend on the modal amplitude $a$, as indicated in \eref{SNMT}.
Normalization with respect to the structure's symmetric and positive definite mass matrix $\mm M$ is assumed so that $\mm\varphi_1^{\mrm H}\mm M\mm\varphi_1 = 1$.
The fundamental Fourier coefficient of the modal vibration, described in terms of the coordinates $\mm q$, is $\hat{\mm q}_1 = \mm\varphi_1 a\ee^{\ii\theta}$.
$\mm q$, $\mm f$ and $\mm M$ refer to the same configuration space.
\\
In the case of base excitation, $\mm f = -\mm M\mm b\ddot q_{\mrm b}$.
Herein, $\mm b$ is Boolean in appropriate coordinates, with entry one if the corresponding coordinate is aligned with the base motion $q_{\mrm b}$ and zero if it is orthogonal.
$\mm q$ is counted relative to the base motion; \ie, $\mm q + \mm b q_{\mrm b}$ is the absolute displacement.
The response coordinate (middle of panel) is defined as $q_{\mrm m} = \mm e^{\mrm T}_{\mrm m}\mm q$.
With this, one can use \eref{SNMT} to derive the frequency response function
\ea{
\frac{\hat q_{\mrm m,1}}{\hat q_{\mrm b,1}} = \frac{(\mm e_{\mrm m}^{\mrm T}\mm\varphi_1(a))\left(\mm\varphi^{\mrm H}_1(a)\mm M\mm b\Omega^2\right)}{-\Omega^2 + 2D(a)\omega(a)\ii\Omega + \omega^2(a)}\fp
\label{eq:qmqb}
}
Herein, $\hat q_{\mrm b,1}$ and $\hat q_{\mrm m,1}$ are the fundamental Fourier coefficients of $q_{\mrm b}$ and $q_{\mrm m}$, respectively.
\eref{qmqb} implies that by keeping $a$, and thus $\left|\hat q_{\mrm m,1}\right| = \left|\mm e_{\mrm m}^{\mrm T}\mm\varphi_1(a)\right|a$, constant over a certain range of the frequency $\Omega$, one obtains quasi-linear behavior.
\\
To achieve a constant fundamental harmonic amplitude $\left|\hat q_{\mrm m,1}\right|$, a feedback controller was used.
To this end, again, synchronous demodulation was used in the present work, to estimate the fundamental harmonic amplitude.
The control error is the deviation to a given response amplitude set value.
A proportional-integral controller was employed, whose output defines the amplitude of the sinusoidal voltage fed to the shaker amplifier.
The modal frequency was simply adopted from the phase resonance condition; thus, RCT and PRT theoretically give the same modal frequency result.
The damping ratio was identified using the common circle fit method.
\\
The RCT was first implemented as proposed in previous works, where the excitation frequency was prescribed and stepped within a range around resonance.
For the given benchmark system in combination with the given excitation system, we were not able to achieve reasonable amplitude control performance $>1~\mrm{Hz}$ (ca. $1\%$) away from resonance.
This impedes the modal testing because the modal frequency is not a priori known, and varies considerably with amplitude, from assembly to assembly and so on.
In order to stay close to resonance, it is thus proposed to carry out the RCT for a fixed phase range (around resonance) rather than a fixed frequency range.
This ensures that the response points are always centered around the modal frequency, independent of its actual value.
Besides overcoming the specific amplitude control problems encountered in the present work, the proposed RCT variant can be interesting because it avoids applying high excitation levels which are typically required further away from resonance.
Such high excitation levels have the potential to induce strongly nonlinear behavior of the excitation system and thus lead to a distorted excitation.
Also, higher excitation levels could lead to excessive heat generation, which may transfer to the structure under test and lead to a more severe time-variability.
Assuming that there is some limitation on the maximum excitation level, higher response levels can be reached with the proposed approach.
The above benefits come at the cost of additional complexity, as one has to use a phase controller (besides the amplitude controller).
As the two controllers operate simultaneously, the feedback loop is more prone to instability, which was in fact observed for the given test rig, as described later.

\subsection{Excitation Controlled Test (ECT)\label{sec:ECT}}
The purpose of the ECT is not to identify modal properties, but to acquire data for cross-validation, as mentioned before.
The method is very similar to the proposed implementation of RCT, where the phase is stepped around resonance, only that the excitation instead of the response level is maintained.

\subsection{Tested configurations and test program\label{sec:protocol}}
Two nominally identical and flat panels were tested.
The photo in \fref{configs}-left, which was obtained by simply placing one panel on the (flat) ground, shows an initial panel curvature.
The flatness deviation is about $1.2~\mathrm{mm}$.
As the curvature is very similar among the tested panels, the curved shape is attributed to the manufacturing process (rolling).
Thanks to the symmetry of the panels, they can be mounted in both ways, so that the initial panel curvature is either \emph{aligned} or \emph{misaligned} with the pillars (\fref{configs}-right).
With the two panels, this gives four configurations:
\begin{enumerate}
    \item[] Configuration 1: Panel 1, Side A (aligned)
    \item[] Configuration 2: Panel 1, Side B (misaligned)
    \item[] Configuration 3: Panel 2, Side A (aligned)
    \item[] Configuration 4: Panel 2, Side B (misaligned)
\end{enumerate}
This terminology is used henceforth for the tested configurations.
\begin{figure}[htb]
 \begin{subfigure}[c]{0.49\textwidth}
     \centering
         \includegraphics[width=1\textwidth]{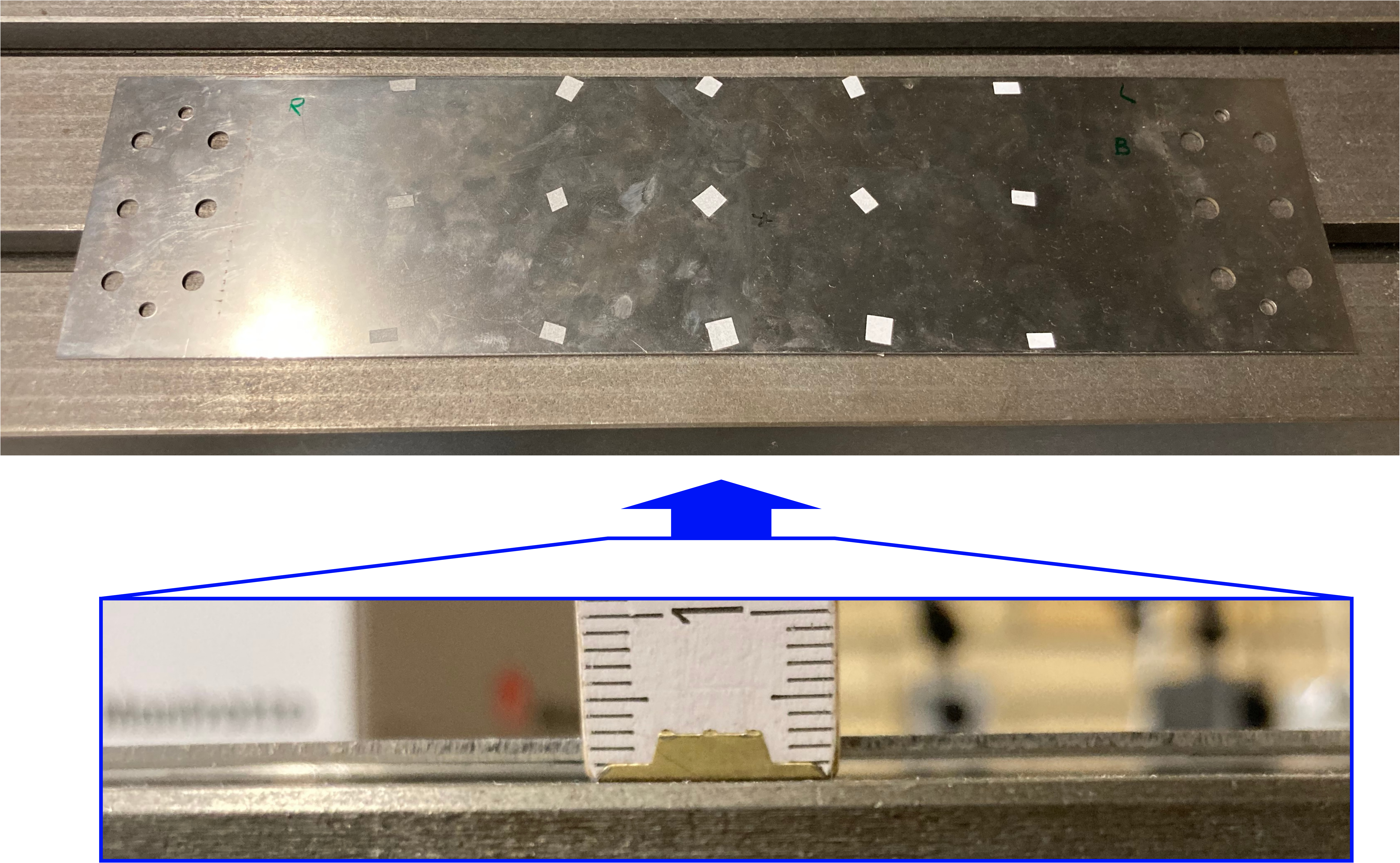}
     \end{subfigure}
     \hfill
\centering
  \begin{subfigure}[c]{0.49\textwidth}
     \centering
         \includegraphics[width=1\textwidth]{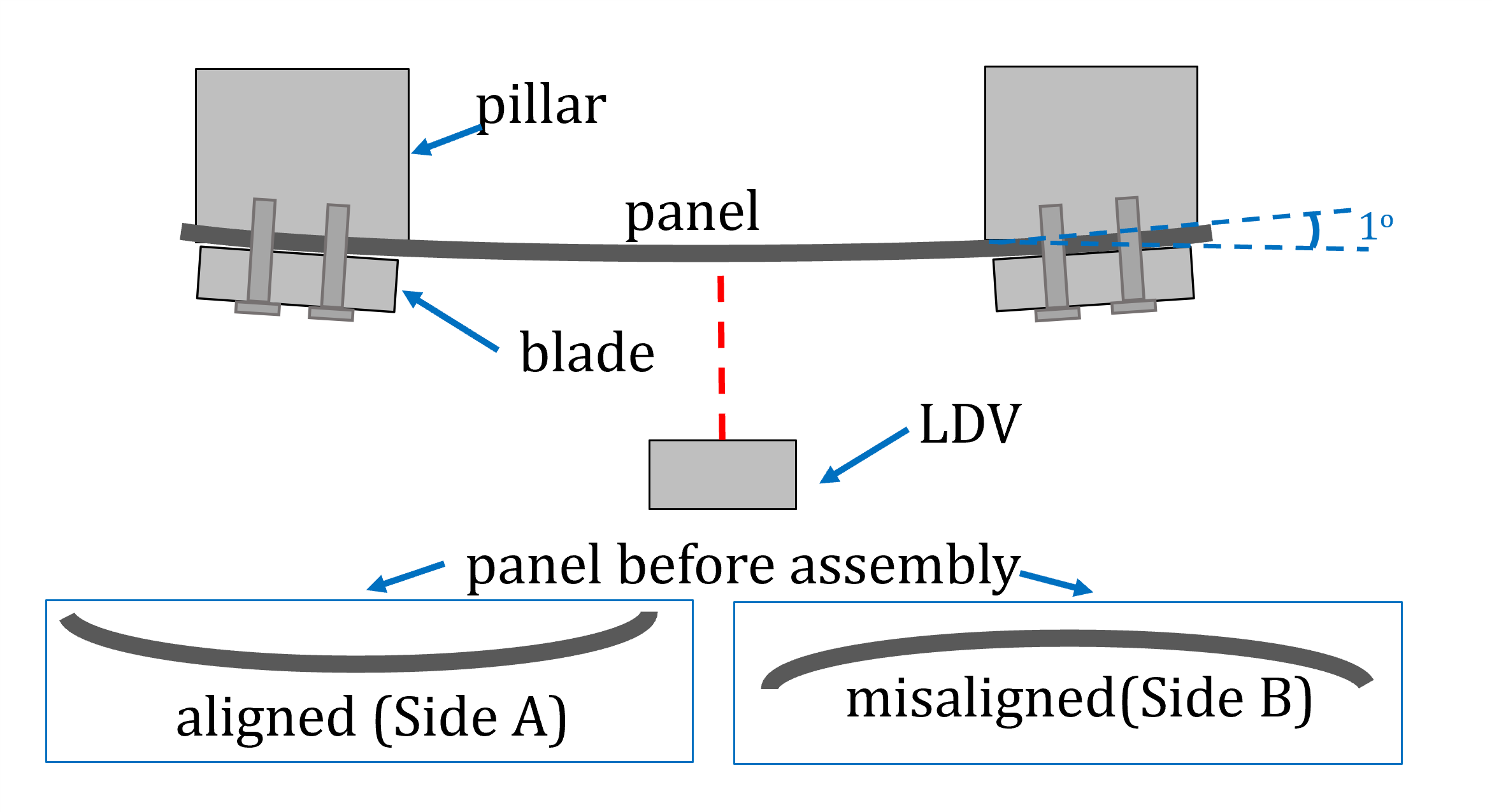}
  \end{subfigure}
       \caption{Tested configurations: Photo of a panel and its initial curvature (left); definition of aligned/misaligned configuration (right).
        }
	\label{fig:configs}
\end{figure}
\\
The test program applied to all four configurations is defined in \fref{sequence}.
Assembly 0 was done only for the characterization of the normal pressure distribution at the four contact interfaces: panel-pillar and panel-blade, on the left and on the right pillar.
The distribution turned out to be quite uniform within the nominal contact area.
More importantly, it was indistinguishable among the configurations, so that a different normal pressure distribution (\eg due to form deviations or waviness) is not regarded as an important source of variability.
The results are shown in \cite{TRCprediction}.
\\
Assembly 1-3 were done for dynamic testing.
A PRT was done in each, whereas ECT only in the first, and RCT only in the last assembly.
Linear modal testing was done before and after each nonlinear test.
For the three types of nonlinear tests, the ranges of the amplitude and the phase, the number of amplitude levels and phase points, and the duration are specified in \tref{par}.
One PRT always included up- and down-stepping of the voltage level.
For configuration 1, more than just one PRT was done per assembly to analyze the unexpected time-variability more thoroughly.
\begin{figure}[htb]
     \centering
         \includegraphics[width=1\textwidth]{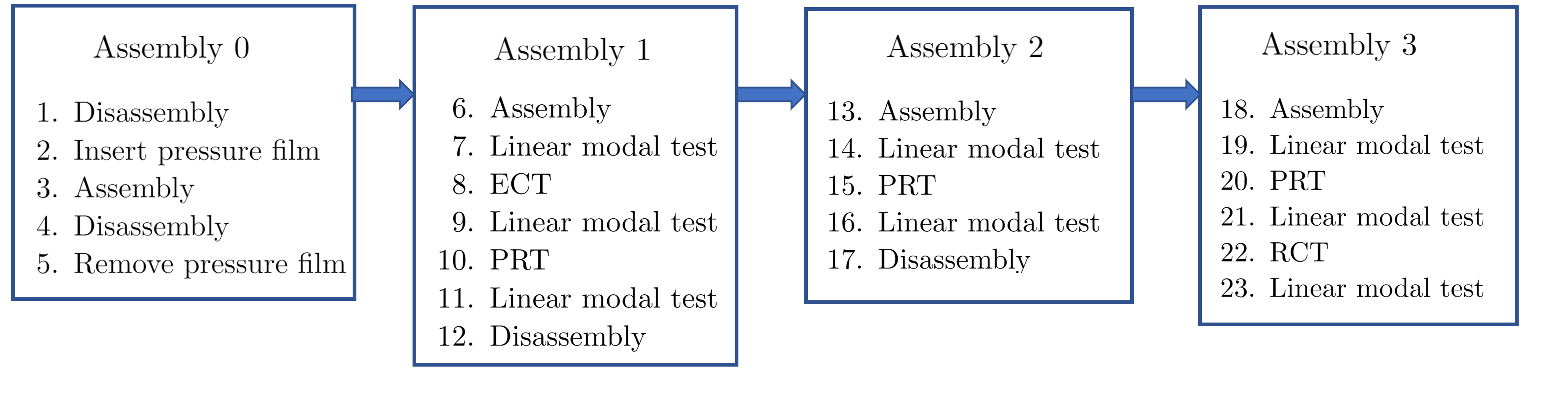}
         \caption{Test program.
         }
         \label{fig:sequence}
\end{figure}
\begin{table}[b]
            \small
			\centering
			\caption{Test parameters}
			\label{tab:par}
			\begin{tabular}{@{}lccc@{}}
				\toprule
				Test & PRT & RCT & ECT \\
                \hline
                Amplitude parameter & shaker voltage & response displacement & base acceleration\\
                Amplitude range & 0.45V-2.6V & 0.2mm-1.4mm & 1m/s$^2$-5m/s$^2$\\
                No. of amplitude levels & 45 & 10 & 5\\
                \hline
                Phase range & $90^\circ$ & $75^\circ$ -$105^\circ$ & $40^\circ$- $140^\circ$\\
                No. of phase points per amplitude level & 2 (up and down) & 12 & 40\\
                \hline
		          Duration per amplitude level & 16 seconds  & 7 minutes &  15 minutes\\
		          Total duration & 24 minutes  & 70 minutes &  75 minutes\\
                \bottomrule
			\end{tabular}
\end{table}
\\
The initial plan was, of course, to span similar response amplitude ranges with all nonlinear tests.
However, due to the poor performance of the combined phase-and-amplitude controller at higher response levels, this was not possible with RCT and ECT.
The corresponding amplitude parameter, response level respectively base acceleration level, had to be restricted more tightly than in the PRT.
A relatively narrow phase range around resonance was deemed sufficient for the RCT.
For the ECT, a wider phase range was preferred to acquire the response both in the nonlinear and in the quasi-linear regime.
As the commercial control hard- and software used for RCT in \cite{Karaagacl.2021,Karaagacl.2022,Koyuncu.2022} was not available, it remains an open question whether the encountered limitations of RCT are specific to the test rig or the implemented, heuristically tuned controllers.

\section{Results\label{sec:results}}
As mentioned before, the focus of the present work was on the system's fundamental bending mode.
First, the results for the linear modal frequency are discussed in \ssref{linearModalParameters}.
Then, the amplitude-dependent modal frequency and damping ratio are addressed in \ssref{amplitudeDependentModalParameters}.
Finally, the modal interaction is analyzed in \ssref{modalInteraction}.

\subsection{Linear modal frequencies\label{sec:linearModalParameters}}
The modal frequencies obtained for the four different configurations using linear modal testing are reported in \fref{f12}.
Time- and reassembly-variability are moderate ($<2\%$) and of similar extent.
Also, the deviation of the mean frequency from panel to panel is considered negligible ($<1\%$); the variability spreads of the corresponding frequencies overlap.
However, there is an appreciable influence of alignment:
The aligned configurations 1 and 3 show softer behavior, while the misaligned configurations 2 and 4 are stiffer.
The mean fundamental modal frequencies deviate by about $6\%$ ($101~\mrm{Hz}$ vs. $107~\mrm{Hz}$), while the mean second modal frequencies deviate by only about $3\%$ ($191~\mrm{Hz}$ vs. $197~\mrm{Hz}$).
No clear trend over the number of tests can be identified.
This also applies to the modal damping ratios (not shown for brevity).
Unlike the modal frequency, the damping ratio does not show any significant effect of the (mis-)alignment.
\begin{figure}[htb]
     \centering
     \begin{subfigure}[c]{0.49\textwidth}
     \centering
         \includegraphics[width=\textwidth]{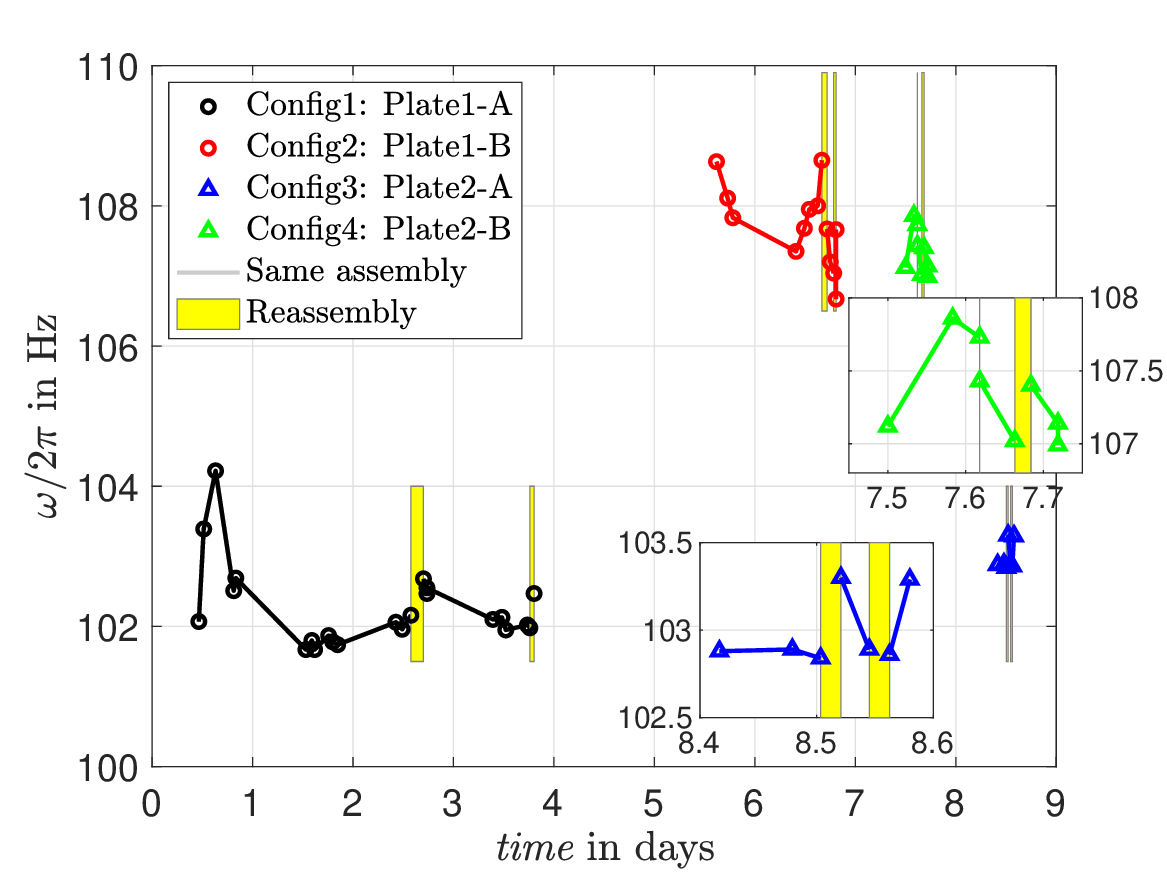}
         \caption{First mode.}
         \label{fig:f1}
      \end{subfigure}
      \hfill
      \centering
      \begin{subfigure}[c]{0.49\textwidth}
          \centering
          \includegraphics[width=\textwidth]{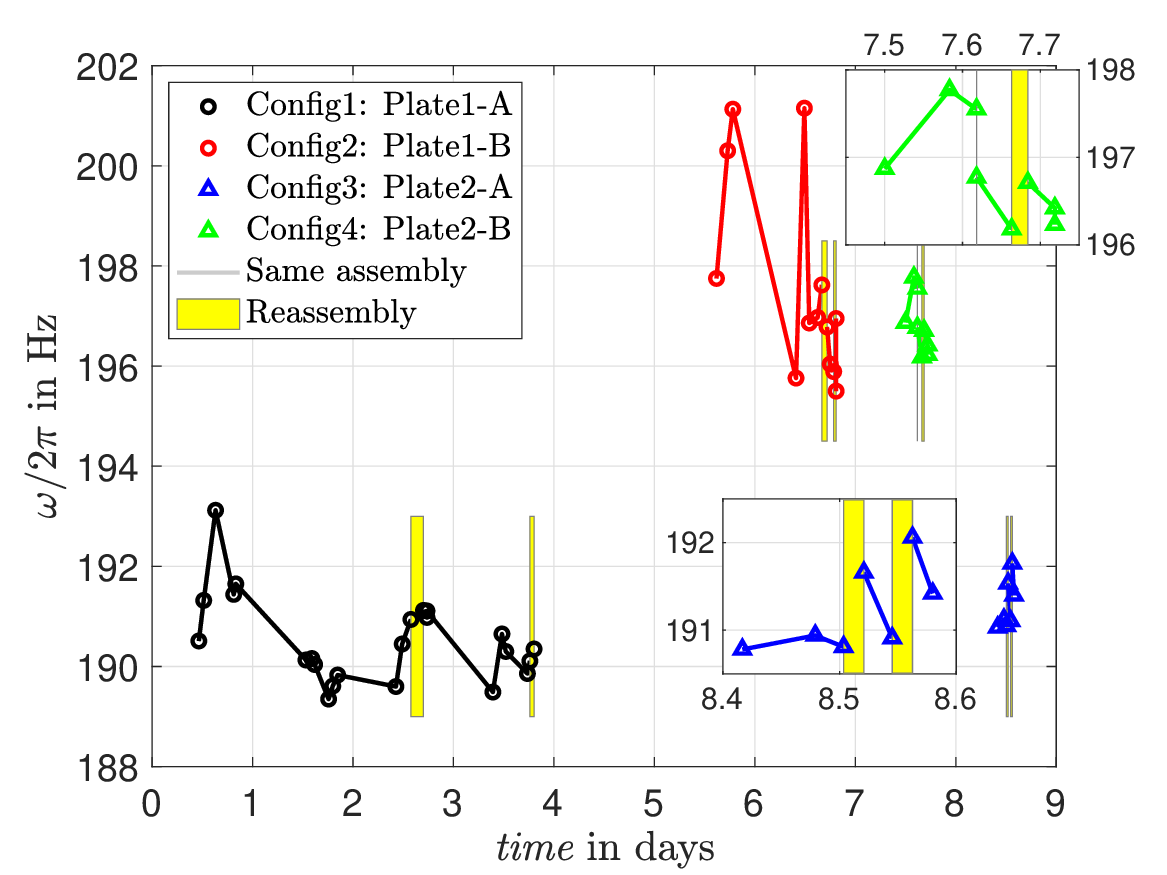}
          \caption{Second mode.}
          \label{fig:f2}
      \end{subfigure}
         \caption{Modal frequencies obtained from linear tests. The markers corresponding to the same assembly are connected by a line and a reassembly is indicated by a yellow square.
         }
         \label{fig:f12}
\end{figure}

\subsection{Amplitude-dependent modal parameters\label{sec:amplitudeDependentModalParameters}}
Next, the time- and reassembly-variability are analyzed for one particular configuration.
Then, the variability from one configuration to another is investigated.
The depicted results are from PRT if not otherwise specified.
Finally, consistency with RCT and cross-validation with ECT results are shown.
\\
As stated before, the response was measured at the panel center, relative to the base motion.
The response displacement was already introduced in \ssref{RCT} as $q_{\mrm m}$.
As amplitude measure, $\hat q_{\mrm m}$, $\sqrt 2$ times the root mean square value of $q_{\mrm m}$ is used throughout the figures in the present work, to be consistent with \cite{TRCprediction}.
Assuming $q_{\mrm m}$ is periodic and given by the Fourier series $q_{\mrm m} = \Re\lbrace \sum_h \hat q_{\mrm m,h}\ee^{\ii h\Omega t}\rbrace$, one can express the amplitude as
\ea{
\hat q_{\mrm m} = \sqrt{\sum\limits_{h=1}^{H} \left|\hat q_{\mrm m,h}\right|^2}\fp \label{eq:amplitude}
}
Here, a finite harmonic order $H$ is assumed; in the present work, the sampling rate allowed to use $H=8$.
Note that a zero-mean displacement is assumed in \eref{amplitude}.
In the experiment, the Fourier coefficients $\hat q_{\mrm m,h}$ were determined by first calculating the Fourier coefficients of the velocity, and then integrating in the frequency domain.

\subsubsection{Time- and reassembly variability}
Configuration 1 underwent the largest number of nonlinear tests.
The results are presented in \fref{remountcon1}.
\begin{figure}[htb]
     \centering
     \begin{subfigure}[c]{0.6\textwidth}
         \centering
     \begin{minipage}{\textwidth}
        \includegraphics[width=1\textwidth]{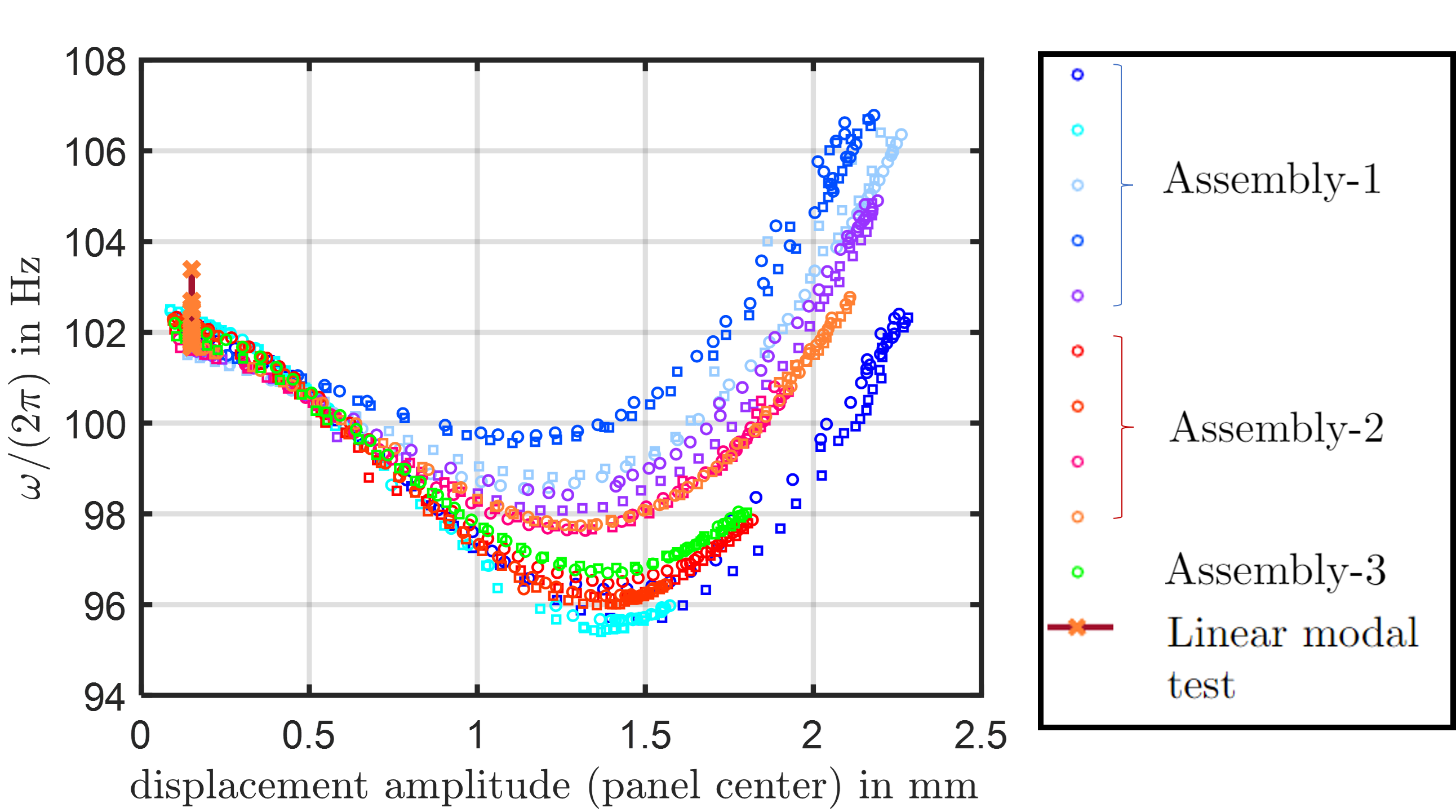}
        \end{minipage}
     \end{subfigure}
     \hfill
     \begin{subfigure}[c]{0.39\textwidth}
         \centering
     \begin{minipage}{\textwidth}
        \includegraphics[width=1\textwidth]{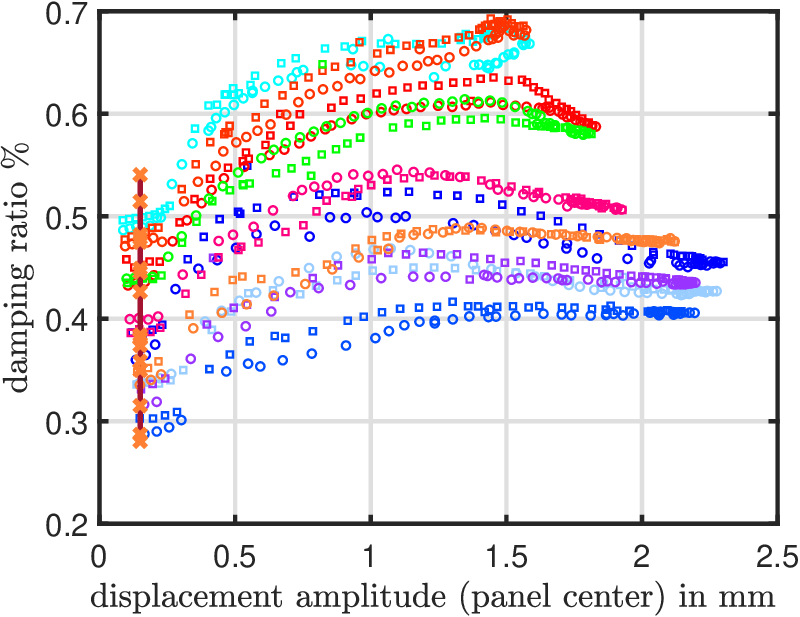}
        \end{minipage}
     \end{subfigure}
     \caption{Time- and reassembly-variability observed for configuration 1: modal frequency vs. amplitude (left); modal damping ratio vs. amplitude (right).
     Circular and square markers indicate up- and down-stepping of the voltage level, respectively, within each PRT.
     }
	\label{fig:remountcon1}
\end{figure}
\\
The tests were carried out over four days.
The same assembly was deliberately tested on more than one day.
The timing of the tests can be inferred from \fref{f12}.
As in the case of the linear modal frequencies, no significant trend over time can be identified.
Such a trend could in principle have been caused by local damage (wear and/or plasticity).
It was generally observed that direct back-to-back tests were very well repeatable.
More variability was found over time.
This is mainly attributed to thermal sensitivity.
The room temperature varied by $5^\circ~\mrm{C}$ over the test program.
The material temperature was generally higher than the room temperature.
Some heat is generated during vibration by dissipation at the contact interfaces and within the bulk.
This explains the slight discrepancy between forward and backward stepping of the voltage amplitude, where the modal frequency tends to slightly decrease over time.
In \fref{remountcon1}, forward stepping is indicated by circular, and backward stepping by square markers.
Also, some heat is transferred from the excitation system, whose temperature varied over the test program.
Interestingly, different maximum response levels were reached for the same maximum voltage amplitude.
This can partially be explained by the damping variability, which should lead to slightly different response levels even for the same base acceleration level.
However, it is concluded that the main reason for the different maximum response levels is that different base acceleration levels were reached for the same voltage level.
This is explained by some time-variability of the excitation system, again, most likely due to thermal effects.
\\
The linear modal test results are also included in \fref{remountcon1}.
They are also plotted at $\sqrt 2$ times the root-mean-square value of $q_{\mrm m}$.
It has to be remarked, however, that $q_{\mrm m}$ has a broad frequency spectrum in the linear tests, as explained in \ssref{linear}.
The identified amplitude is most likely an upper estimate of the effective amplitude in the resonant mode, as some energy is contained in higher-order modes.
Overall, the nonlinear modal test results are consistent with linear ones at low amplitude.
This holds both for the modal frequency and the damping ratio.
\\
The modal frequency shows a well-repeatable softening-hardening trend.
The initial softening is not surprising for a curved plate, since bending can induce membrane compression, leading to a considerable reduction of the bending stiffness.
Also, the hardening trend can be explained by the dominance of membrane stretching at higher displacement amplitudes.
The damping ratio initially increases, then saturates or even decreases slightly in some cases.
The initial increase is typical for partial/micro-slip in bolted connections.
The saturation/decrease is typical for gross slip.
However, it is unlikely that gross slip was reached in the contact interfaces due to the relatively high normal pressure and the dense arrangement of the bolts.
It seems likely that this trend was due to the mutual interaction with the nonlinear bending-stretching coupling, but this is not fully understood yet.
In general, the higher variability of the damping ratio, compared to that of the modal frequency, is typical for jointed structures.

\subsubsection{Panel-to-panel and alignment variability} 
The amplitude-dependent modal properties are shown for the four different configurations in \fref{configvar}. The markers depict the mean of all tests for each configuration, and the shaded region represents the spread of the data within each configuration. The data were truncated to the overlap of the individual displacement ranges for each configuration.
As in the linear case, there is a clear difference between the aligned configurations, 1 and 3, and the misaligned configurations, 2 and 4:
Besides a shift along the frequency axis, the transition from hardening to softening seems to be shifted slightly towards higher amplitudes for the misaligned configurations.
Taking into account the time- and reassembly-variability, indicated by the shaded regions, the difference between the two configurations 1 and 3, or 2 and 4 is almost negligible, with regard to both, frequency and damping.
\begin{figure}[htb]
     \centering
     \begin{subfigure}[c]{0.49\textwidth}
         \centering
        \includegraphics[width=1\textwidth]{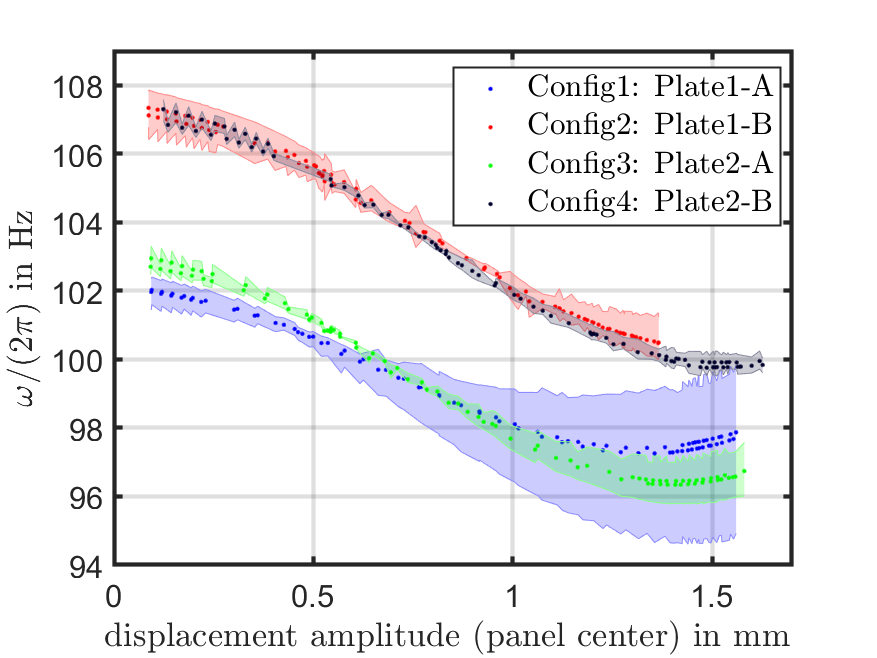}
     \end{subfigure}
     \hfill
     \begin{subfigure}[c]{0.49\textwidth}
         \centering
        \includegraphics[width=1\textwidth]{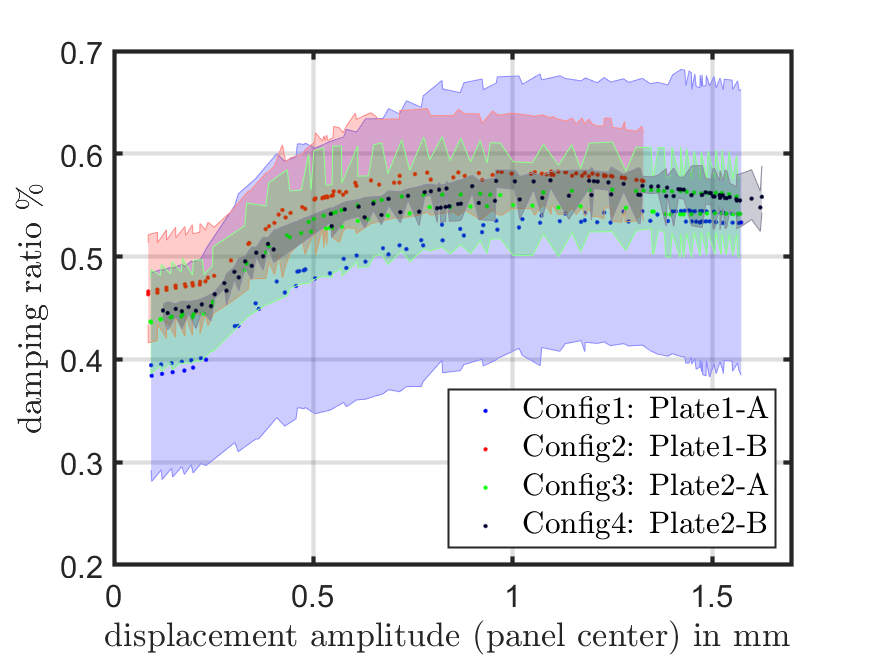}
     \end{subfigure}
       \caption{Configuration-variability: modal frequency vs. amplitude (left); modal damping ratio vs. amplitude (right).
       }
	\label{fig:configvar}
\end{figure}

\subsubsection{Consistency of phase resonance with response controlled test}
Due to the aforementioned instability of the feedback loop, the RCT had to be limited to $1.4~\mrm{mm}$ maximum response amplitude.
Additionally, severe steady-state control errors were observed beyond about $0.6~\mrm{mm}$ response amplitude.
A correlation was observed between the occurrence of severe control errors and the presence of a strong second harmonic in the acquired velocity signals.
What aspect of the interplay between the combined controllers, the synchronous demodulation and the nonlinear structure under test causes this phenomenon is not fully understood yet.
The resulting control errors are analyzed and discussed in \aref{circlefit}.
The consistency check should be limited to response points without severe control errors.
To this end, the following thresholds were set for accepting a given steady-state response:
\begin{enumerate}
\item The deviation of the response amplitude is $<2~\%$,
\item the standard deviation of the frequency is $<0.2~\mrm{Hz}$, and
\item the standard deviation of phase is $<2.5^\circ$.
\end{enumerate}
In the PRT, for reference, the phase error was $<2^\circ$, and the frequency had a standard deviation $<0.35~\mrm{Hz}$. With this, the amplitude tolerance is larger than the nominal amplitude step, and the phase tolerance is equal to the nominal phase step (\tref{par}).
The frequency tolerance seems somewhat large given that the entire frequency span covered for the phase range $75^\circ-105^\circ$ is $<0.5~\mrm{Hz}$, typically about $0.3~\mrm{Hz}$.
It should be remarked that there is no rigorous theory or comparable experience available that would inform the decision on the tolerances.
Also, as one can see in \aref{circlefit}, practically no acceptable points would be left for a much tighter frequency tolerance.
In any case, by rejecting the response points that exceed the above specified tolerances, the RCT results depicted in \fref{compRCTresults} are obtained.
These are compared to the PRT results acquired during assembly 3 (\fref{sequence}), \ie, immediately before the RCT.
This way, assembly-variability is eliminated, and the effect of time-variability is minimized.
 \begin{figure}[h!]
     \centering
     \begin{subfigure}[c]{0.49\textwidth}
         \centering
        \includegraphics[width=1\textwidth]{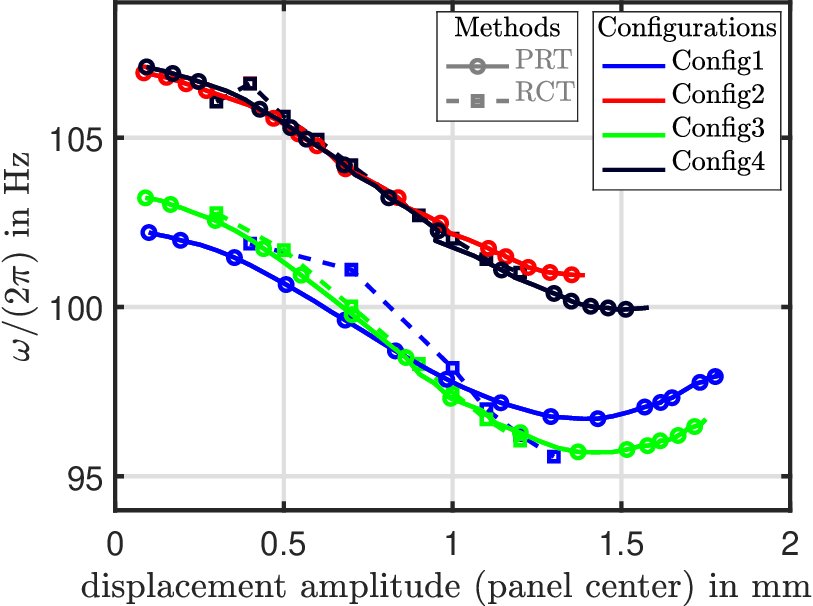}
     \end{subfigure}
     \hfill
     \begin{subfigure}[c]{0.49\textwidth}
         \centering
        \includegraphics[width=1\textwidth]{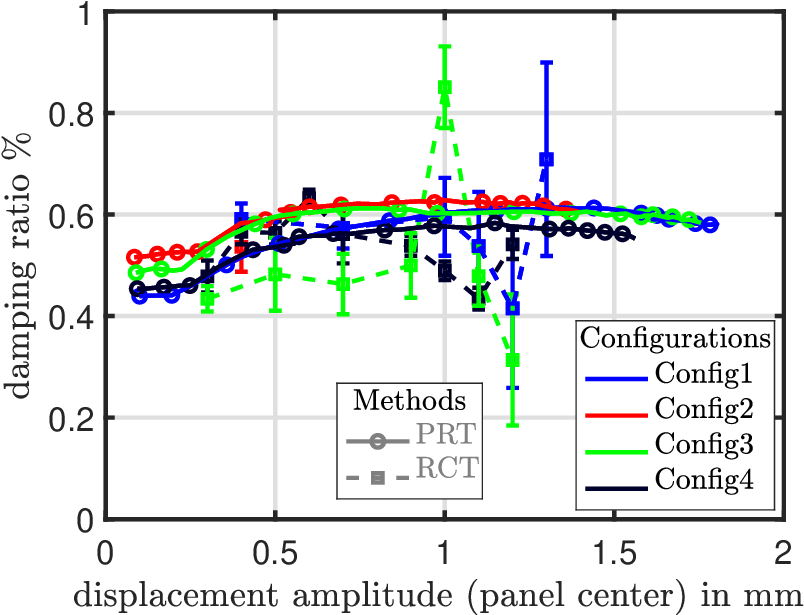}
     \end{subfigure}
       \caption{Consistency of RCT and PRT results: modal frequency vs. amplitude (left); modal damping ratio vs. amplitude (right).}
	\label{fig:compRCTresults}
\end{figure}
\\
Recall that the modal frequency was identified from the phase resonance condition, also for RCT.
Thus, the deviations in the modal frequency are attributed to the remaining control errors and some system-inherent variability.
\\
The modal damping ratios were identified using the circle fit method.
Here, any pair of points of different phase can be used to evaluate the damping ratio, as illustrated in \aref{circlefit}.
All possible combinations were considered to obtain a set of estimates.
The error bars in \fref{compRCTresults}-right indicate the mean, the minimum and the maximum values of this set.
The PRT results are largely within the error bars of the RCT results.
For configuration 3 and 4, the rejection of points with pronounced amplitude variation left only 4-6 points for the circle fit at the second- and third-highest amplitude points.
Thus, one should have relatively low confidence in those results.
\\
In summary, it can be said that the poor control performance induces considerable uncertainty with regard to most results obtained with the RCT method in the present work.
Focusing on the reasonably well-controlled results, and taking into account the uncertainty spread, there is good qualitative and quantitative agreement between PRT and RCT results.
This applies both with respect to the modal frequency and the damping ratio, and the quantitative agreement applies to both the mean value and the evolution with the amplitude.

\subsubsection{Cross-validation between phase resonance and excitation controlled test}
In this subsection, a cross-validation is carried out by predicting the near-resonant frequency response curve for fixed excitation level, using the nonlinear modal oscillator model identified via PRT, and comparing the results with the direct measurements (ECT).
In principle, any configuration could be used for this.
However, configuration 1 and 3 show substantial internal resonance phenomena, as shown later in \ssref{modalInteraction}, so that single-nonlinear-mode theory cannot be expected to give reasonable results.
In contrast, such phenomena remain weak in configurations 2 and 4, so that the nonlinear modal oscillator model should be valid.
Among those two, configuration 4 was selected because it showed the least variation in the linear modal frequency.
The ECT results were acquired in assembly 1, see \fref{sequence}, and the PRT results obtained in this assembly were used to feed the modal oscillator model.
The results of the cross-validation are shown in \fref{FRC}.
Only response points meeting the control error tolerances, as defined for the RCT results, are depicted.
\begin{figure}[h!]
 \centering
 \begin{subfigure}{0.49\textwidth}
     \includegraphics[width=\textwidth]{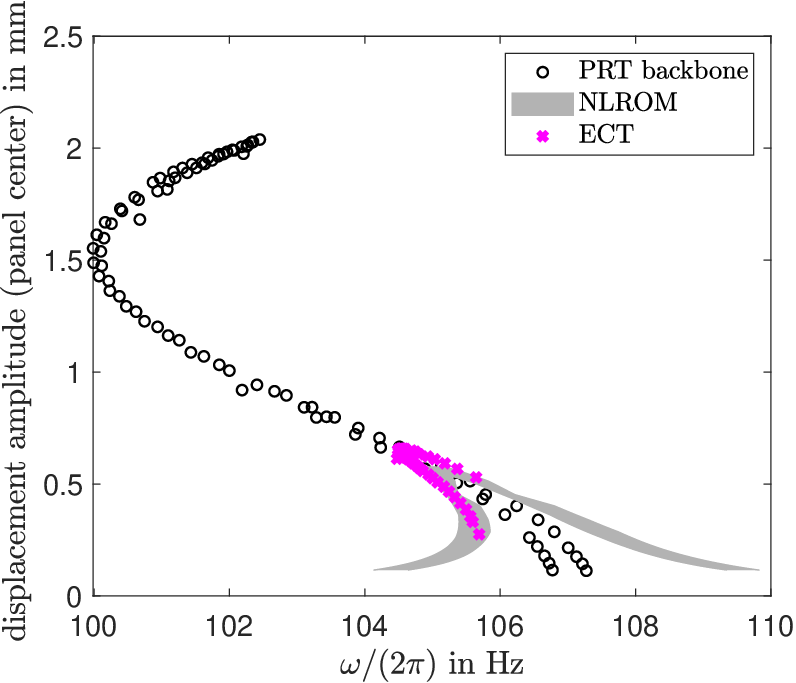}
     \caption{Excitation level 2 m/$s^2$}
 \end{subfigure}
 \hfill
 \begin{subfigure}{0.49\textwidth}
     \includegraphics[width=\textwidth]{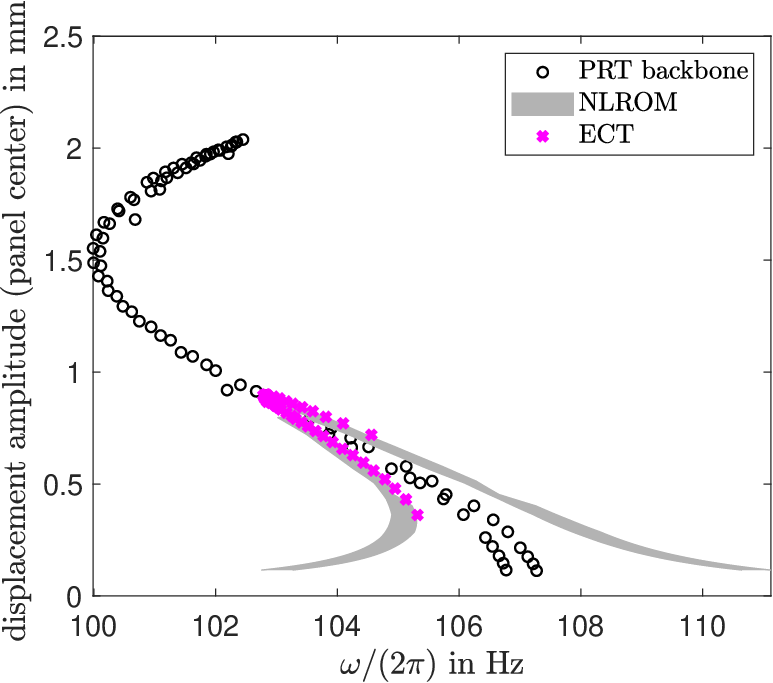}
     \caption{Excitation level 3 m/$s^2$}
 \end{subfigure}
  \hfill
 \begin{subfigure}{0.49\textwidth}
     \includegraphics[width=\textwidth]{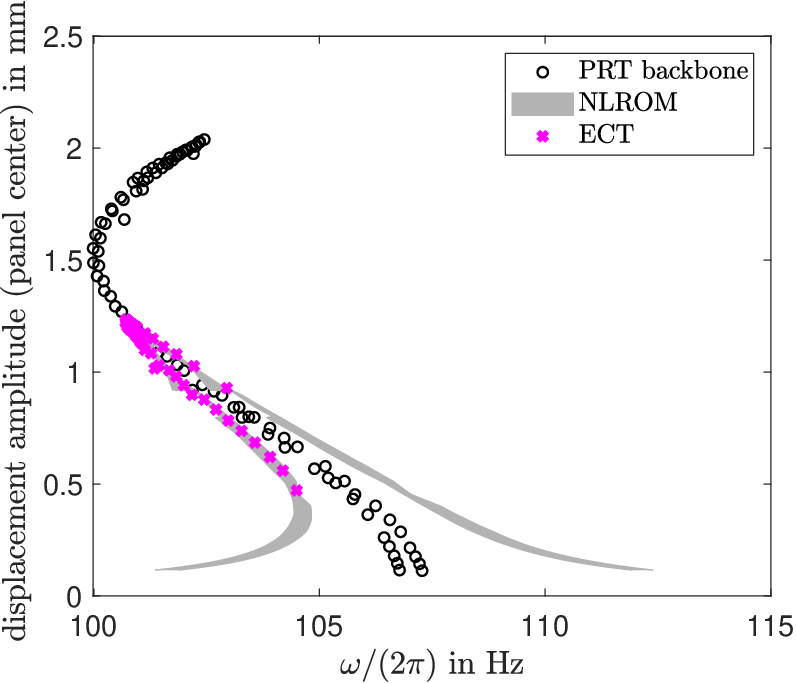}
     \caption{Excitation level 4 m/$s^2$}
 \end{subfigure}
 \hfill
 \begin{subfigure}{0.49\textwidth}
     \includegraphics[width=\textwidth]{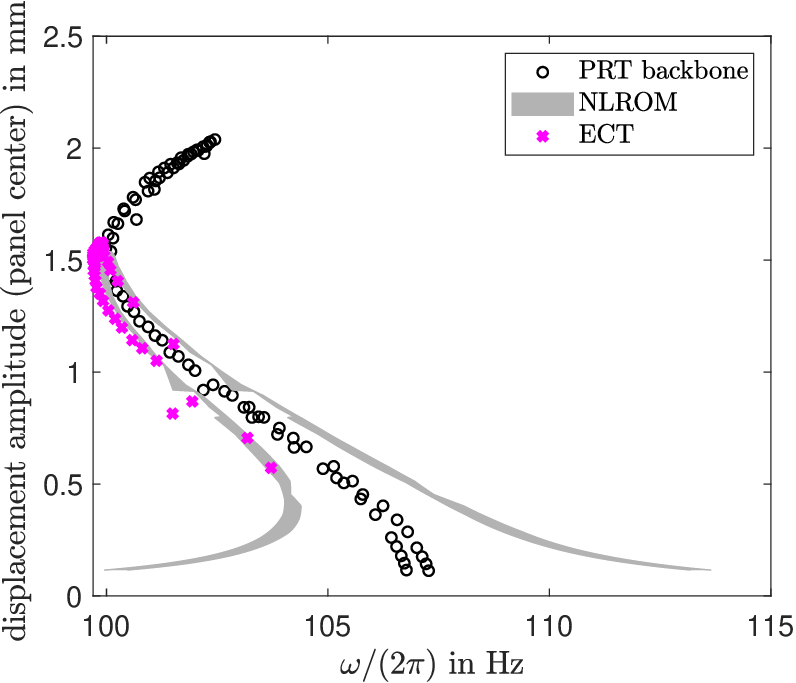}
     \caption{Excitation level 5 m/$s^2$}
 \end{subfigure}
         \caption{Cross-validation between phase resonance and excitation controlled test, configuration 4.
         }
    \label{fig:FRC}
\end{figure}
\\
The technique developed in \cite{Ferhatoglu.2023} was used to obtain the frequency response bounds in accordance with \eref{SNMT}, taking into account the slight difference between up and down stepping results obtained in the PRT.
The ECT results largely fall inside these bounds.
Also, the resonance peak obtained by ECT coincides in good approximation with the (phase-resonant) backbone.
As expected, the frequency response curve is strongly bent towards the left, and features an overhanging branch, already for the lowest tested base acceleration amplitude of $2~\mrm{m}/\mrm{s}^2$.
At the highest tested excitation level, the frequency response curve starts bending towards the right.
Unfortunately, higher excitation levels could not be tested due to the aforementioned control issues.

\subsection{Modal interaction\label{sec:modalInteraction}}
The two lowest-frequency modes are the first bending and the first torsion mode.
Their deflection shapes are shown in \fref{modeShapes}.
Those two modes are not very far from a 1:2 internal resonance condition (\ssref{linearModalParameters}).
Further, the first torsion mode has a damping ratio of only $<0.04\%$, which is about one order of magnitude less than that of the first bending mode.
This condition is likely to amplify possible internal resonance phenomena.
\fig[h!]{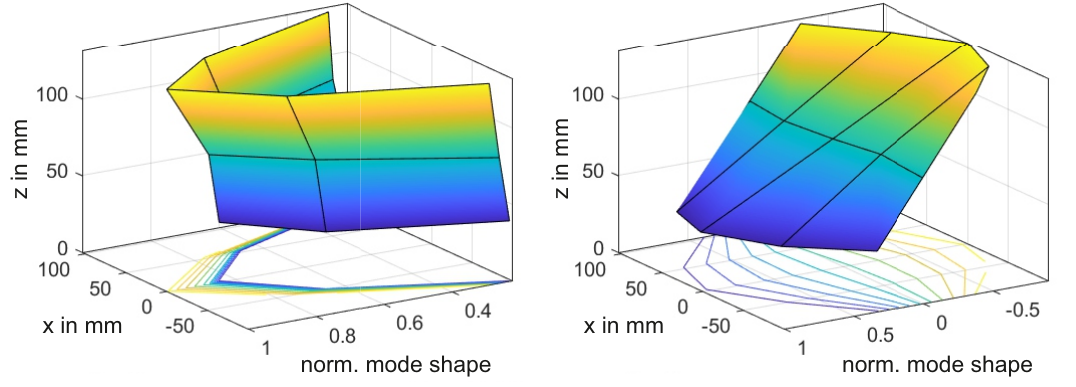}{Low-amplitude mode shapes: (left) first bending, (middle) first torsion.}{.75}
%
\\
To determine whether or not a strong modal interaction takes place, the individual modal and harmonic contributions to the vibration were estimated.
For this, it is useful to consider the decomposition of the period-averaged mechanical energy,
\ea{
E_{\mrm{mech}} = \sum_{m} \sum_{h} \underbrace{\frac14 \left[\left(h\Omega\right)^2+\omega_m^2\right]\left|\hat\eta_m(h)\right|^2}_{E(m,h)}\fp \label{eq:energy}
}
Herein, $\omega_m$ is the linear frequency of mode $m$, and $\hat\eta_m(h)$ is the $h$-th complex Fourier coefficient of the $m$-th modal coordinate $\eta_m$ (assumed periodic here).
\eref{energy} assumes that the modal deflection shapes are mass-normalized.
From the steady-state velocity data acquired during PRT, the relevant contributions $E(m,h)$ were estimated.
%
\fig[h!]{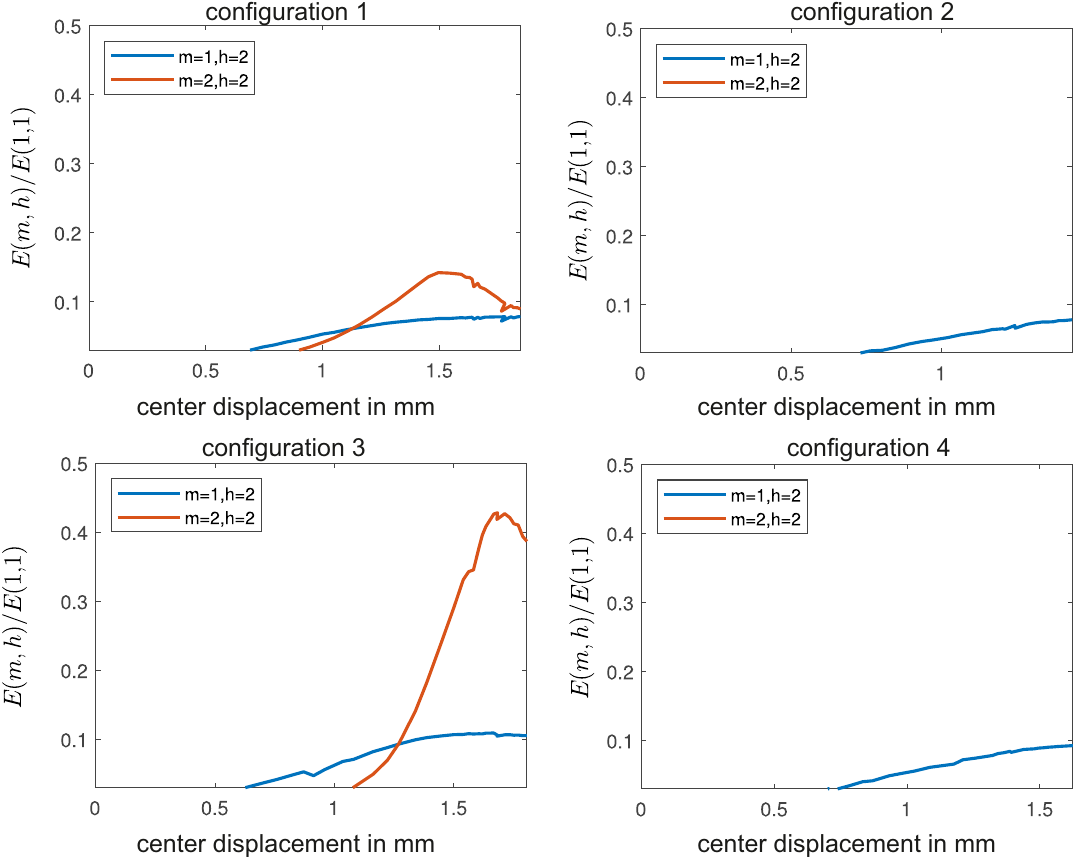}{Dominant modal and harmonic contributions to vibration energy for the 4 configurations.}{1.0}
\\
In the linear resonant case, the only non-negligible contribution should be $E(1,1)$.
\fref{modalEnergies} shows the contribution of the second harmonic of mode 1 and the second harmonic of mode 2, relative to $E(1,1)$, for all four configurations.
The remaining harmonic and modal contributions never exceeded $3\%$ and are thus below the depicted range.
In the aligned configurations 1 and 3, clear internal resonance phenomena are observed:
$E(2,2)$ reaches a peak at $15~\%$ and $45\%$ energy fraction, respectively.
In the misaligned configurations 2 and 4, in contrast, the contribution of the second mode is negligible.
This is attributed to the fact that the frequency ratio is closer to 1:2 in the aligned configuration.
More specifically, the frequency ratio is $1.89$ in the aligned and $1.84$ in the misaligned configuration (\ssref{linearModalParameters}).
Still, an appreciable second harmonic contribution of the resonant mode, $E(1,2)$, is observed.
The second harmonic is typical for initially curved plates.
Minimal models include quadratic (besides cubic) polynomial terms, see \eg \cite{Camier.2009}.
Assuming a perfectly symmetric plate subjected to an ideal base excitation, one would still expect that the torsion mode does not respond.
However, the mounting of the panel on the pillars destroys the symmetry of the setup.
The pillars undergo a slight cantilever-type deflection.
Indeed, a small elastic deformation of about $4~\%$ along the pillar length was observed.
As a result, the bending mode also exhibits larger deflections towards the upper edge of the panel (\fref{modeShapes}-left).
Apparently, this is sufficient to permit the observed nonlinear interaction with the first torsion mode.

\section{Conclusions\label{sec:concl}}
Thanks to a thorough test design and several means of validation, results of very high confidence have been obtained for the amplitude-dependent frequency and damping ratio of the TRC benchmark system's fundamental mode.
This is a crucial prerequisite for the further development of prediction approaches.
Research on prediction method development will likely take several years, given the fact that open questions arise already for systems that have only geometric or contact nonlinearity, and the benchmark system has shed light into interesting mutual interactions.
An important part of the present work was to disambiguate and quantify sources of variability.
The sensitivity to inevitable thermal effects and shape deviations was found to affect both the bending-stretching and the contact behavior.
The vast majority of state-of-the-art methods is not able to account for both nonlinear behavior and uncertainty, and this is regarded as one of the most imminent shortcomings.
Finally, the poor performance of the phase controller, in particular when combined with an amplitude controller, shows the strong need for research towards systematic and robust control design.

\section{Declaration of Competing Interest}
The authors declare that they have no known competing financial interests or personal relationships that
could have appeared to influence the work reported in this paper.

\section*{Acknowledgements}
M. Krack is grateful for the funding received by the Deutsche Forschungsgemeinschaft (DFG, German Research Foundation) [Project 450056469, 495957501].\\
This work presents results of the Tribomechadynamics Research Camp (TRC). The authors thank MTU Aero Engines AG for sponsoring the TRC 2022.\\
The authors are grateful to Maximilian W. Beck for designing the benchmark system.\\
The support from NSF grant No: 1847130 is appreciated by A. Bhattu.\\
S. Hermann is grateful for the funding received by the EIPHI Graduate School, ANR-17-EURE-0002.\\
N. Jamia gratefully acknowledges the support of the Engineering and Physical Sciences Research Council through the award of the Programme Grant “Digital Twins for Improved Dynamic Design”, grant number EP/R006768/1.

\appendix
\setcounter{figure}{0}
\setcounter{table}{0}

\section{Control performance in the response controlled test and resulting uncertainty\label{asec:circlefit}}
In the response controlled tests, the response displacement of the plate center, $q_{\mrm m}$, was set as target and the base displacement, $q_{\mrm b}$ was determined accordingly. The measurement data obtained for different amplitudes in configuration 4 is shown as an example in \fref{appendixRCT1}~a and~b respectively. Amplitude and phase of the frequency response function are shown in \fref{appendixRCT1}~c and d to illustrate that a resonance peak is obtained for each displacement amplitude.
\begin{figure}[h!]
        \centering
       \includegraphics[trim = 65 0 70 0, clip,width=1\textwidth]{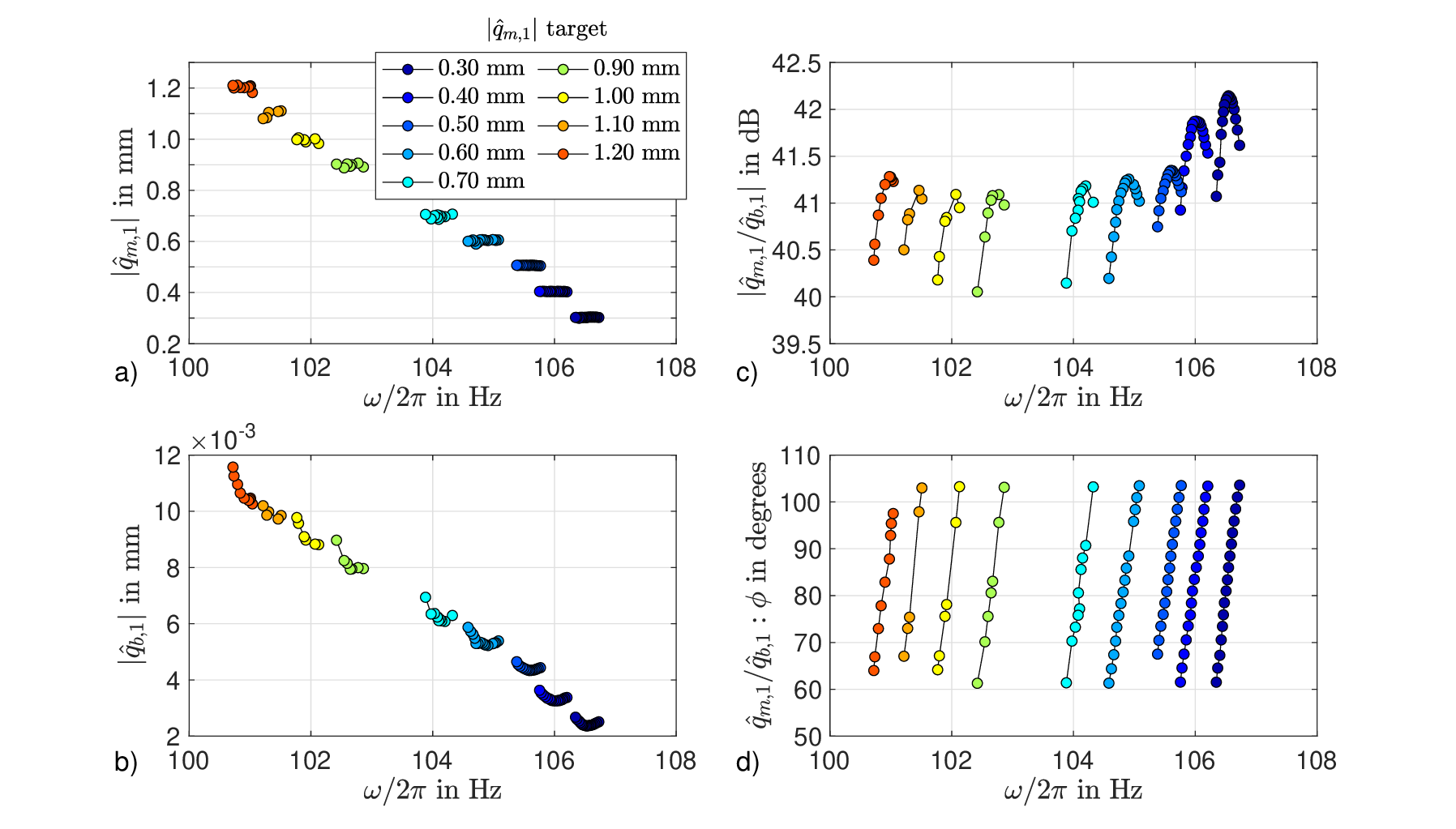}
        \caption{Measurement results obtained for configuration 4: panel center amplitude (a), base amplitude (b), amplitude (c) and phase (d) of the frequency response function.}
	\label{fig:appendixRCT1}
\end{figure}

\begin{figure}[h!]
        \centering
       \includegraphics[trim = 65 0 70 0, clip,width=1\textwidth]{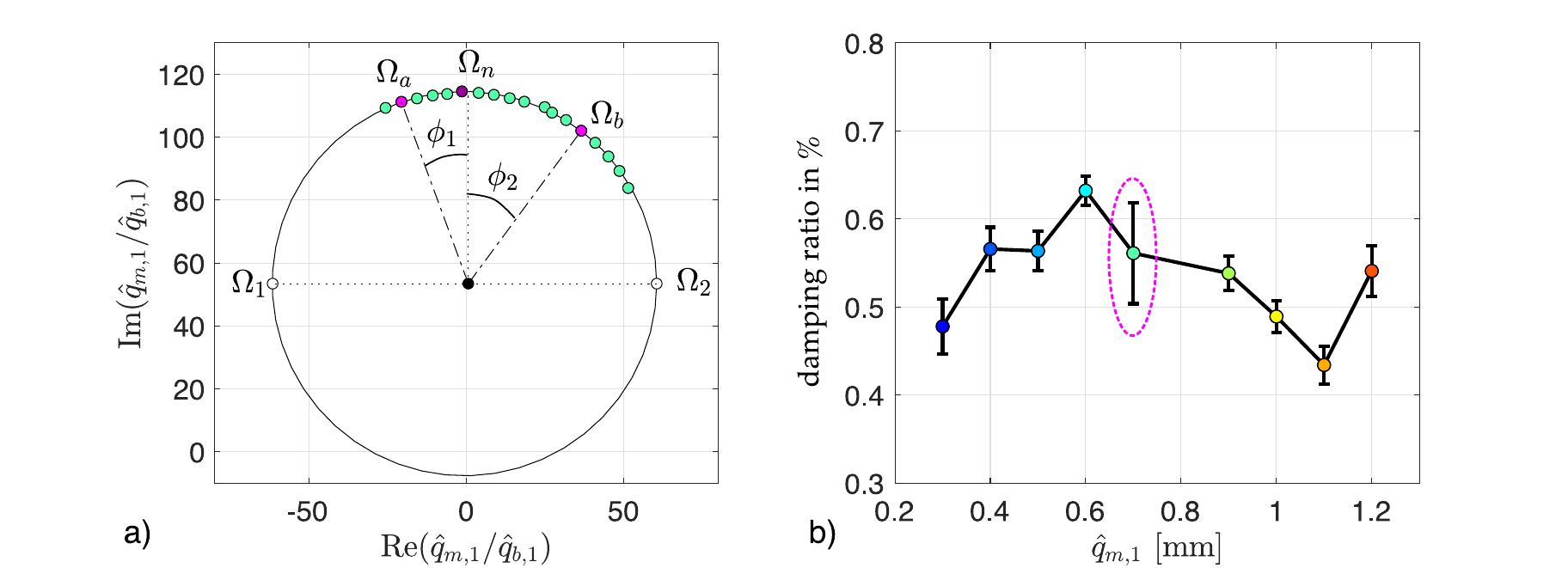}
        \caption{Measurement values obtained for configuration 4: frequency response function for $\hat{q}_{m,1} =$~\SI{0.7}{\milli \meter} represented in a Nyquist diagram along with fitted circle (a) and damping ratio with highlighted amplitude $\hat{q}_{m,1} =$~\SI{0.7}{\milli \meter} (b).}
	\label{fig:appendixRCT2}
\end{figure}

When the frequency response function is represented in a Nyquist diagram, its real and imaginary parts form a circle (\cf \fref{appendixRCT2}~a). To obtain the damping ratio, a circle is fitted through the measurement points in a first step. The resonance is located at the point with the highest imaginary part and the half-power points are located where the circle has the highest and lowest real part respectively (frequencies: $\Omega_1$ and $\Omega_2$).
\textbf{A commonly used method to determine the resonance frequency is using the spacing between the points on the arc of the Nyquist diagram: the maximum spacing for equal frequency increments occurs at the resonance. In our experimental approach, however, we obtain results for equal phase increments which leads to equally spaced points on the arc (cf. Fig.~\ref{fig:appendixRCT2}a, varying distances are due to measurement and control uncertainty). Therefore, the measurement value with the smallest real and highest imaginary part was considered as resonant, with the corresponding frequency $\Omega_n$ in the present work. }
The damping ratio can be obtained from
\begin{equation}
D= \frac{{\Omega_b}^2-{\Omega_a}^2}{{2 \,\Omega_n}^2} \,  \left(\tan\frac{\phi_1}{2} + \tan \frac{\phi_2}{2}\right)^{-1},
\label{eq:rctDamping}
\end{equation}
where $\Omega_a$ and $\Omega_b$ are the frequencies of arbitrary measurement values above and below the resonance respectively. $\phi_1$ and $\phi_2$ describe the angles between the two chosen points and the resonant point (\cf \fref{appendixRCT2}~a).
To include the measurement uncertainty of the points in the results, the damping was calculated for all combinations of points above and below the resonance. Finally, mean and standard deviation were computed (\cf \fref{appendixRCT2}~b).

As previously mentioned, the RCT results suffer from high control errors.
\fref{appendixRCT3}~a shows the results obtained for the damping ratio prior to filtering the control errors. In several cases, a significant error between the response target and the resulting measurement occurred. An example is shown in \fref{appendixRCT3}~b. To exclude the outliers from the evaluation, only results with an amplitude deviation $<2\%$ were considered in the evaluation (\cf \fref{appendixRCT1}~a). The outliers appear more often for response amplitudes between \SI{0.6}{\milli \meter} and \SI{1.0}{\milli \meter} than outside of this interval and the effect is therefore considered to be related to the control difficulties during the modal interaction. Another difficulty during the tests was the control of the phase difference. The deviation of the phase from the target increased with the response amplitude, as illustrated in \fref{appendixRCT3}~c. The phase fluctuation affects the frequency which also shows a higher standard deviation with increasing response amplitude. The effect is not fully understood yet, but the influence of this control error on the circle fit method is limited by excluding results with a phase and frequency deviation higher than \SI{3.5}{\degree} and \SI{0.2}{\hertz} respectively. The filtered results are shown in \fref{compRCTresults}.
\begin{figure}[h!]
        \centering
       \includegraphics[trim = 65 0 70 0, clip,width=1\textwidth]{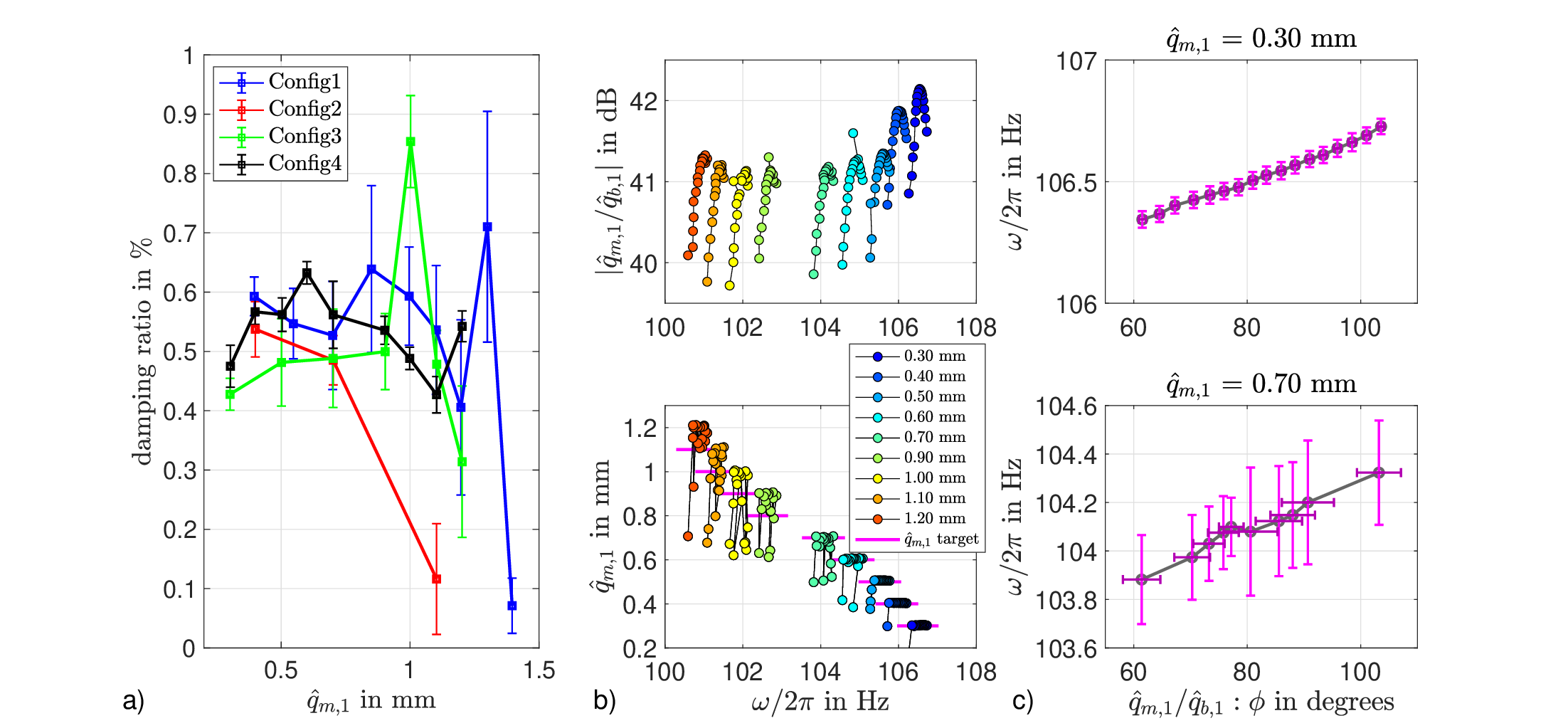}
        \caption{Results obtained prior to filtering of badly controlled steady states: Damping ratio of all configurations (a), frequency response function (top) and response displacement (bottom) of configuration 4 for all tested response displacements (b) and phase-frequency plots showing the mean and standard deviation of both quantities, response displacements of \SI{0.3}{\milli \meter} (top) and \SI{0.7}{\milli \meter} (bottom) (c).}
	\label{fig:appendixRCT3}
\end{figure}

\end{document}